\documentclass[
reprint,
superscriptaddress,
amsmath,amssymb,
aps,
prx,
longbibliography,
showkeys,
]{revtex4-2}

\usepackage{graphicx}
\usepackage{dcolumn}
\usepackage{bm}

\usepackage{tcolorbox}
\usepackage{booktabs,tabularx,array}
\usepackage{amsmath}
\usepackage[colorlinks=true,linkcolor=blue,citecolor=blue,urlcolor=blue]{hyperref}

\begin{document}

\preprint{APS/123-QED}

\title{Conservative Charge and Current Deposition on Nonuniform 3D Cylindrical PIC Meshes with Residual Self-Field Diagnostics
}%

\author{Zhe Liu}
\author{Zhijun Zhou}

\author{Yinjian Zhao}
 \email[Contact author: ]{zhaoyinjian@hit.edu.cn}
 
\affiliation{School of Energy Science and Engineering, Harbin Institute of Technology, Harbin 150001, People's Republic of China}

\date{\today}

\begin{abstract}
Particle-in-cell simulations on nonuniform cylindrical meshes require deposition schemes that respect cylindrical metrics while controlling numerical self-fields. This work develops cylindrical-volume-weighted charge and current deposition using nodal control volumes and swept-volume factors on logically structured stretched grids. 
Uniform-density and controlled-transport tests demonstrate accurate charge recovery, well-controlled current-density errors, and continuity residuals substantially lower than the corresponding current-density errors.
Single-particle diagnostics show that charge-transport consistency alone does not ensure self-field cancellation; face-centered electric fields give the smallest residual, 
whereas cell-centered and shifted cell-centered layouts produce larger residuals.
\end{abstract}

\maketitle

\section{Introduction}

Particle-in-cell (PIC) methods provide a standard kinetic framework for plasma simulations by coupling charged particles and electromagnetic fields through deposition and interpolation. In many devices, such as plasma thrusters, Z-pinches, and dense plasma-focus discharges\cite{taccogna2023plasma,adam2004study,taccogna2005plasma,angus2024implicit,schmidt2012fully,krishnan2012dense,zhao2026review}, both the geometry and key dynamics, including radial compression, axial acceleration, and azimuthal motion, are more naturally represented in cylindrical formulations. Nonuniform meshes are also widely used to resolve localized gradients near electrodes, sheaths, plasma boundaries, or other localized high-gradient regions without refining the entire domain. These choices improve physical resolution and computational efficiency, but they also make particle--field coupling more delicate than in uniform Cartesian grids, because the cylindrical volume element, $r\,dr\,d\phi\,dz$, and stretched mesh spacings modify the discrete measures used in deposition, field gradients, and particle gather operations\cite{ruyten1993density,araki2014cell,stanier2022curvilinear}.

In a PIC algorithm, deposition refers to the procedure of mapping particle properties, such as charge and current, onto the computational mesh to form the discrete source terms for the field solver. The accuracy and consistency of this step are therefore essential for reliable particle--field coupling. On a uniform Cartesian mesh, charge-conserving current deposition has been extensively studied, and schemes such as those of Villasenor--Buneman and Esirkepov distribute the swept charge during particle motion to satisfy the discrete continuity equation \cite{villasenor1992rigorous,esirkepov2001exact,li2005volume}. In cylindrical coordinates, this Cartesian simplification no longer holds because the control volume varies with radius; accordingly, density-conserving weighting and corrected-volume deposition schemes have been developed to account for the radial metric factor \cite{ruyten1993density,verboncoeur2001symmetric,araki2014cell}. The issue becomes more restrictive on nonuniform cylindrical meshes, where both the radial metric factor and the stretched mesh spacings must be included in the deposition procedure. For charge deposition, the particle charge should be normalized by the appropriate nodal control volume rather than by a Cartesian cell volume. For current deposition, the charge swept during particle motion must be distributed consistently with the finite-volume metric factors and the discrete continuity equation. Otherwise, apparently reasonable local assignment weights may still produce geometry-induced bias in the deposited density and current.

\begin{figure*}[htbp]
    \centering
    \includegraphics[width=\linewidth]{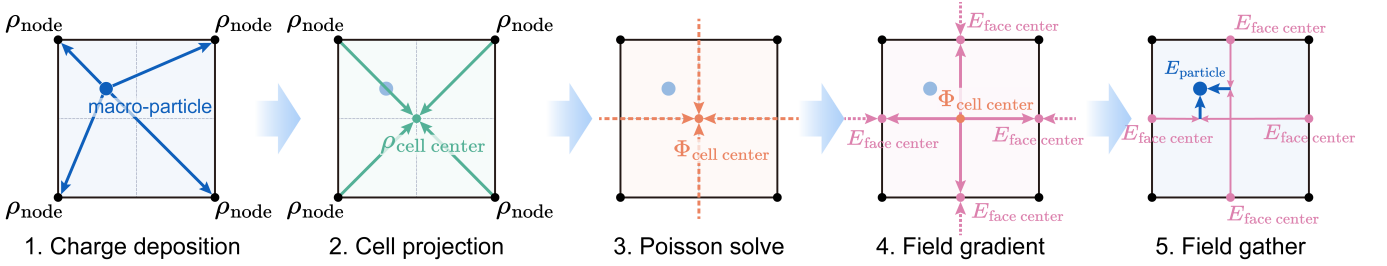}
    \caption{Particle--field coupling workflow considered in this work. Charge and current deposition are derived and verified for charge-transport consistency, while the residual self-field diagnostic is performed in the electrostatic Poisson--gradient--gather branch. The workflow is shown on a 2D cell for clarity, whereas the deposition formulation and numerical tests are performed on 3D cylindrical meshes.}
    \label{fig:deposition flowchart}
\end{figure*}

After deposition, the grid-based sources enter the field solver\cite{bao2025accelerated}. In the electrostatic branch considered for the self-field diagnostic, the deposited charge density is used in a finite-volume Poisson solve, the electric field is obtained from the potential gradient, and the field is then gathered back to particle positions. A numerical self-field arises when the charge assignment, field solve, and gather interpolation are not mutually compatible, leaving a particle to sample a finite contribution from its own deposited source \cite{hockney2021computer,lira2020self}. In ideal uniform Cartesian PIC formulations, translational symmetry and matched charge-assignment and field-interpolation operators can cancel or strongly suppress this self-force \cite{hockney2021computer}. Geometric consistency in deposition alone is therefore not sufficient to guarantee self-field cancellation. On nonuniform cylindrical grids, this cancellation becomes more difficult because the deposition weights, control volumes, finite-volume field solve, field-component placement, and gather interpolation all depend on the local radius and mesh spacing. Although self-force subtraction and compatible geometric formulations have been developed for more general meshes and coupling schemes \cite{lira2020self,moon2015exact}, these studies indicate that self-field cancellation must be assessed at the level of the full scatter--solve--gradient--gather chain, rather than from deposition consistency alone. The particle--field coupling workflow considered in this work is summarized in Fig.~\ref{fig:deposition flowchart}.

The present work is closely related to the conservative cylindrical
deposition method of Zhao \emph{et al.}~\cite{zhao2023rigorously},
which extended the Villasenor--Buneman swept-volume
current-deposition scheme~\cite{villasenor1992rigorous} from
Cartesian coordinates to three-dimensional cylindrical PIC on Yee
grids. That work established charge--current consistency for
first-order deposition on uniform cylindrical cells and removed edge
errors through corrected cylindrical volume factors. However, it did
not address locally stretched cylindrical meshes, nor did it examine
the residual self-field produced by the subsequent field solve and
particle gather. The present study extends the deposition formulation
to logically structured nonuniform cylindrical finite-volume meshes
and further evaluates how the complete Poisson--gradient--gather chain
affects self-field cancellation.

For nonuniform cylindrical finite-volume PIC schemes, the main difficulty is not only to construct geometrically consistent deposition formulas, but also to determine whether these formulas remain compatible with the subsequent field solve and gather operations. Existing studies on cylindrical or curvilinear PIC deposition have mainly addressed density-conserving weighting, corrected control volumes, charge-conserving current deposition, and compatibility with the discrete continuity equation \cite{ruyten1993density,verboncoeur2001symmetric,araki2014cell,zhao2023rigorously,stanier2022curvilinear,xiao2018structure,xiao2021explicit}. However, the residual self-field generated by the complete scatter--solve--gradient--gather chain on nonuniform cylindrical finite-volume meshes has been less directly characterized.
 This leaves an important gap: a deposition scheme may be conservative and geometrically consistent, but still produce a finite self-force if the field placement, finite-volume discretization, and gather interpolation are not mutually compatible. Such a residual force can contaminate particle acceleration, distort local particle dynamics, and introduce artificial energy errors in simulations where field gradients and particle trajectories are important. Although field placement and scatter--gather compatibility have been recognized as important issues in PIC coupling schemes \cite{moon2015exact,zoni2022hybrid,shapoval2024pseudospectral}, a systematic comparison of face-centered, cell-centered, and shifted electric-field layouts for self-field cancellation on stretched cylindrical finite-volume meshes remains limited. It is also not clear whether the resulting residual self-field is mainly caused by cylindrical curvature or by mesh nonuniformity. A systematic diagnostic is therefore needed to separate charge and current conservation from self-field cancellation in nonuniform cylindrical PIC formulations.

In this work, we examine charge and current density deposition on nonuniform cylindrical grids and characterize the associated residual self-field. Cylindrical-volume-weighted charge and current deposition formulas are constructed for logically structured nonuniform meshes and verified using a uniform particle distribution and a controlled particle-transport test. These tests evaluate density recovery, current-density accuracy, and consistency with the discrete continuity equation. The analysis is then extended to the residual self-field by applying a single-particle diagnostic to face-centered, cell-centered, and shifted cell-centered field-placement layouts. A Cartesian-limit cylindrical test is further performed to separate the effect of grid stretching from that of cylindrical curvature. The study demonstrates that charge-transport consistency and self-field cancellation are related but distinct properties, and that the residual self-field is controlled by the compatibility of the full scatter--solve--gradient--gather chain.

\section{Method}
\label{sec:method}

\subsection{Geometric measures on nonuniform cylindrical grids}
\label{subsec:geometric_measures_nonuniform_cylindrical}

A logically structured cylindrical grid is used in the present work.
The grid nodes are denoted by $(r_i,\phi_j,z_k)$, where
$i=0,\ldots,N_r$, $j=0,\ldots,N_\phi$, and
$k=0,\ldots,N_z$. The cell indices are
$i=0,\ldots,N_r-1$, $j=0,\ldots,N_\phi-1$, and
$k=0,\ldots,N_z-1$. The mesh spacing is allowed to be
nonuniform in all three directions, with
\[
\Delta r_i=r_{i+1}-r_i,\quad
\Delta \phi_j=\phi_{j+1}-\phi_j,\quad
\Delta z_k=z_{k+1}-z_k .
\]
The corresponding cell is
$
C_{i,j,k}
=
[r_i,r_{i+1})\times[\phi_j,\phi_{j+1})\times[z_k,z_{k+1}) .
$
The azimuthal direction is treated as periodic unless otherwise stated. In
the periodic case, $\phi_{N_\phi}$ is identified with
$\phi_0+2\pi$, and azimuthal indices are interpreted modulo $N_\phi$.

In cylindrical coordinates, the physical volume element is
$dV=r\,dr\,d\phi\,dz$. Therefore, the exact volume of cell
$C_{i,j,k}$ is
\begin{equation}
\Omega_{i,j,k}
=
\frac{1}{2}
\left(r_{i+1}^{2}-r_i^{2}\right)
\Delta\phi_j\Delta z_k .
\label{eq:cell_volume_cylindrical}
\end{equation}
Here $\Omega_{i,j,k}$ is used for cell-based geometric quantities, such
as finite-volume source terms.
Since the cell sizes in the radial, azimuthal, and axial directions are generally different on a nonuniform cylindrical mesh, a scalar measure of the local cell size is introduced for normalization purposes. We define this effective local mesh size as the side length of a cube with the same volume,
\begin{equation}
h_{\mathrm{eff},i,j,k}
=
\Omega_{i,j,k}^{1/3}.
\end{equation}

In the present formulation, charge density is stored at grid nodes. Each
node is therefore associated with a finite-volume control volume. To avoid
confusion with the cell volume $\Omega_{i,j,k}$, the nodal control volume
is denoted by
\begin{equation}
V^{\mathrm{node}}_{i,j,k}
=
V_i^r V_j^\phi V_k^z ,
\label{eq:nodal_control_volume}
\end{equation}
where $V_i^r$, $V_j^\phi$, and $V_k^z$ are the radial, azimuthal, and
axial nodal measures, respectively.

The cylindrical Jacobian is included in the radial nodal measure. For an
interior radial node $1\le i\le N_r-1$, the radial measure is obtained by
integrating the first-order nodal basis function with the metric factor
$r$:
\begin{equation}
\begin{aligned}
V_i^r
&=
L_i^{r,-}+L_i^{r,+},\\
L_i^{r,-}
&=
\int_{r_{i-1}}^{r_i}
r\frac{r-r_{i-1}}{r_i-r_{i-1}}\,dr
=
\frac{(r_i-r_{i-1})(2r_i+r_{i-1})}{6},\\
L_i^{r,+}
&=
\int_{r_i}^{r_{i+1}}
r\frac{r_{i+1}-r}{r_{i+1}-r_i}\,dr
=
\frac{(r_{i+1}-r_i)(2r_i+r_{i+1})}{6}.
\end{aligned}
\label{eq:radial_nodal_measure_integral}
\end{equation}
Equivalently,
\begin{equation}
V_i^r
=
\frac{
(r_{i+1}-r_{i-1})(r_{i+1}+r_i+r_{i-1})
}{6},
\quad
1\le i\le N_r-1 .
\label{eq:radial_nodal_measure_compact}
\end{equation}

At a radial boundary, the missing half interval is omitted. Thus, at the
inner radial boundary,
\begin{equation}
V_0^r
=
\frac{(r_1-r_0)(2r_0+r_1)}{6},
\end{equation}
which reduces to $V_0^r=r_1^2/6$ when the cylindrical axis
$r_0=0$ is included. At the outer radial boundary,
\begin{equation}
V_{N_r}^r
=
\frac{(r_{N_r}-r_{N_r-1})(2r_{N_r}+r_{N_r-1})}{6}.
\end{equation}

For interior or periodic azimuthal nodes, the azimuthal nodal measure is
\begin{equation}
V_j^\phi
=
\frac{\Delta\phi_{j-1}+\Delta\phi_j}{2}.
\end{equation}
For interior axial nodes, the axial nodal measure is
\begin{equation}
V_k^z
=
\frac{\Delta z_{k-1}+\Delta z_k}{2}.
\end{equation}
At physical azimuthal or axial boundaries, the corresponding one-sided
half-width is used.

The quantities defined in this subsection provide the geometric
normalization used in the charge and current deposition formulas below.
In particular, $V^{\mathrm{node}}_{i,j,k}$ converts deposited nodal charge
into charge density, while the local mesh spacings and cylindrical radial
factors enter the current-density deposition.

\subsection{Charge deposition to nodal control volumes}
\label{subsec:charge_density_deposition_nonuniform_cylindrical}

Using the geometric measures defined in
Sec.~\ref{subsec:geometric_measures_nonuniform_cylindrical}, consider a
particle $p$ with charge $q_p$, numerical weight $w_p$, and position
$(r_p,\phi_p,z_p)$. Let
\begin{equation}
Q_p=q_p w_p
\end{equation}
be the represented particle charge. Suppose that the particle lies in cell
$C_{i,j,k}$, namely
\[
(r_p,\phi_p,z_p)
\in
[r_i,r_{i+1})\times[\phi_j,\phi_{j+1})\times[z_k,z_{k+1}) .
\]
Near the periodic azimuthal boundary, the angular coordinate is unwrapped
locally before evaluating the interpolation weights.

A first-order cloud-in-cell (CIC) shape function is used in the logical
cell. The one-dimensional weights are
\begin{equation}
\begin{aligned}
S_i^r
&=
\frac{r_{i+1}-r_p}{\Delta r_i},
&
S_{i+1}^r
&=
\frac{r_p-r_i}{\Delta r_i},\\
S_j^\phi
&=
\frac{\phi_{j+1}-\phi_p}{\Delta\phi_j},
&
S_{j+1}^\phi
&=
\frac{\phi_p-\phi_j}{\Delta\phi_j},\\
S_k^z
&=
\frac{z_{k+1}-z_p}{\Delta z_k},
&
S_{k+1}^z
&=
\frac{z_p-z_k}{\Delta z_k}.
\end{aligned}
\label{eq:cic_weights_cylindrical}
\end{equation}
These weights satisfy
\begin{equation}
S_i^r+S_{i+1}^r=1,\qquad
S_j^\phi+S_{j+1}^\phi=1,\qquad
S_k^z+S_{k+1}^z=1 .
\end{equation}

The three-dimensional CIC weight for the surrounding node
$(i+a,j+b,k+c)$, with $a,b,c\in\{0,1\}$, is
\begin{equation}
S^{abc}_{i,j,k}
=
S_{i+a}^r S_{j+b}^\phi S_{k+c}^z .
\end{equation}
The deposited nodal charge contribution from particle $p$ is then
\begin{equation}
\Delta Q^{\mathrm{node},(p)}_{i+a,j+b,k+c}
=
Q_p S_{i+a}^r S_{j+b}^\phi S_{k+c}^z,
\quad
a,b,c\in\{0,1\}.
\label{eq:nodal_charge_deposition}
\end{equation}
In the periodic azimuthal direction, the index $j+1=N_\phi$ is mapped
back to $0$ after the local weights are evaluated.

After contributions from all particles have been accumulated, the nodal
charge density is obtained by normalizing the nodal charge with the nodal
control volume:
\begin{equation}
\rho_{i,j,k}
=
\frac{
Q^{\mathrm{node}}_{i,j,k}
}{
V^{\mathrm{node}}_{i,j,k}
}.
\label{eq:nodal_charge_density}
\end{equation}
Equivalently, the density increment associated with a single particle is
\begin{equation}
\Delta \rho^{(p)}_{i+a,j+b,k+c}
=
\frac{
q_p w_p
S_{i+a}^r S_{j+b}^\phi S_{k+c}^z
}{
V^{\mathrm{node}}_{i+a,j+b,k+c}
},
\quad
a,b,c\in\{0,1\}.
\label{eq:nodal_charge_density_increment}
\end{equation}

If the cylindrical axis is included, all azimuthal copies of an axis node
represent the same physical point. In that case, the accumulated axis
charge is first combined over the azimuthal direction, and the axis density
is evaluated as
\begin{equation}
\rho_{0,j,k}
=
\frac{
\sum_m Q^{\mathrm{node}}_{0,m,k}
}{
V_0^r V_k^z \sum_m V_m^\phi
},
\quad
\forall j .
\label{eq:axis_charge_density_average}
\end{equation}
For a uniform azimuthal mesh, this reduces to an arithmetic azimuthal
average at the axis.

The total charge density is obtained by summing over all particles. If
number density is required instead of charge density, $Q_p=q_pw_p$ is
replaced by $w_p$, or equivalently the factor $q_p$ is omitted.

\subsection{Current deposition from swept-volume factors}
\label{subsec:current_density_deposition_nonuniform_cylindrical}

\begin{figure}
    \centering
    \includegraphics[width=0.8\linewidth]{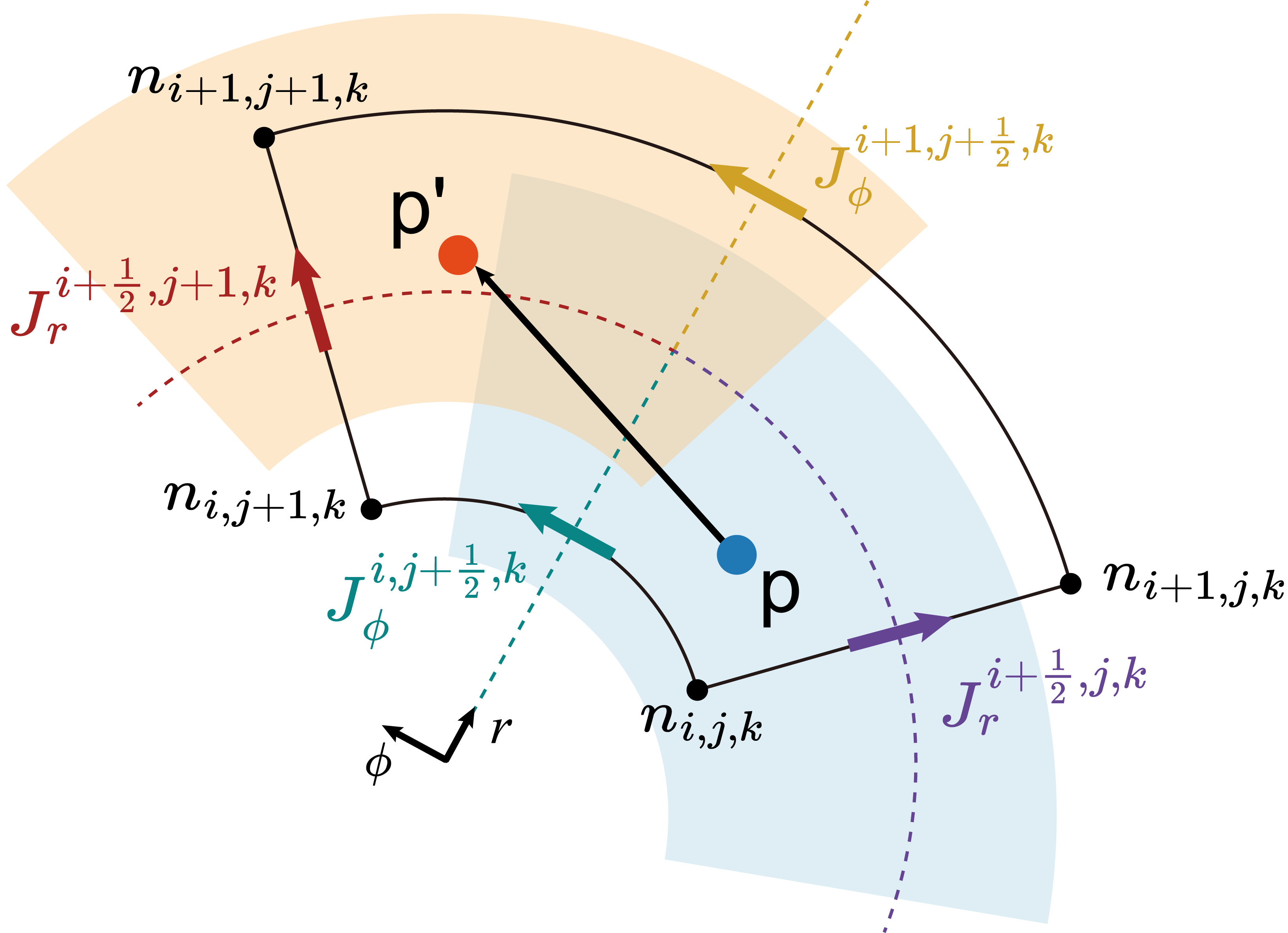}
    \caption{Particle trajectory segment for current deposition.}
    \label{fig:particle motion}
\end{figure}

The current density is obtained from the charge swept by a particle during one time step. The same nonuniform cylindrical mesh and nodal measures introduced in Sec.~\ref{subsec:geometric_measures_nonuniform_cylindrical} are used in the following derivation. The local trajectory segment used for swept-volume current deposition is illustrated in Fig.~\ref{fig:particle motion}.

Consider a particle moving from $(r_p,\phi_p,z_p)$ to $(r_p',\phi_p',z_p')$ over a time interval $\Delta t$. The corresponding velocities are $v_r=(r_p'-r_p)/\Delta t$, $v_\phi=(\phi_p'-\phi_p)/\Delta t$, and $v_z=(z_p'-z_p)/\Delta t$, where $v_\phi$ denotes the angular velocity. The trajectory is approximated as linear within one deposition interval:
\begin{equation}
\begin{aligned}
r_p(s) &= r_p + v_r s,\\
\phi_p(s) &= \phi_p + v_\phi s,\\
z_p(s) &= z_p + v_z s,
\end{aligned}
\qquad 0 \le s \le \Delta t .
\end{equation}
Equivalently, the finite displacements are denoted by $\delta r=v_r\Delta t$, $\delta\phi=v_\phi\Delta t$, and $\delta z=v_z\Delta t$. The expressions below are written for a trajectory segment fully contained in one cell. If a particle crosses a cell boundary during a time step, 
the trajectory is decomposed into cell-contained subsegments following the
standard Villasenor--Buneman splitting procedure~\cite{villasenor1992rigorous},
and the same deposition formula is applied to each subsegment.

For a segment located in cell $(i,j,k)$, define the local distances
\begin{equation}
\begin{aligned}
r^+ &= r_{i+1}-r_p,        & r^- &= r_p-r_i,\\
\phi^+ &= \phi_{j+1}-\phi_p, & \phi^- &= \phi_p-\phi_j,\\
z^+ &= z_{k+1}-z_p,        & z^- &= z_p-z_k .
\end{aligned}
\end{equation}
The radial midpoint of the particle trajectory is $\widetilde r_p=r_p+\delta r/2$, and the radial position of the radial-current face is $r_{i+1/2}=(r_i+r_{i+1})/2$.

The transverse kernels associated with the radial current are
\begin{equation}
\begin{aligned}
A_r^{00}
&=
\phi^+ z^+
-\frac{\phi^+\delta z}{2}
-\frac{z^+\delta\phi}{2}
+\frac{\delta\phi\delta z}{3},\\
A_r^{10}
&=
\phi^- z^+
-\frac{\phi^-\delta z}{2}
+\frac{z^+\delta\phi}{2}
-\frac{\delta\phi\delta z}{3},\\
A_r^{01}
&=
\phi^+ z^-
+\frac{\phi^+\delta z}{2}
-\frac{z^-\delta\phi}{2}
-\frac{\delta\phi\delta z}{3},\\
A_r^{11}
&=
\phi^- z^-
+\frac{\phi^-\delta z}{2}
+\frac{z^-\delta\phi}{2}
+\frac{\delta\phi\delta z}{3}.
\end{aligned}
\label{eq:Ar}
\end{equation}
Here the two superscript indices denote the azimuthal and axial directions, respectively.

For the azimuthal current, the corresponding transverse kernels are
\begin{equation}
\begin{aligned}
A_\phi^{00}
&=
r^+ z^+
-\frac{r^+\delta z}{2}
-\frac{z^+\delta r}{2}
+\frac{\delta r\delta z}{3},\\
A_\phi^{10}
&=
r^- z^+
-\frac{r^-\delta z}{2}
+\frac{z^+\delta r}{2}
-\frac{\delta r\delta z}{3},\\
A_\phi^{01}
&=
r^+ z^-
+\frac{r^+\delta z}{2}
-\frac{z^-\delta r}{2}
-\frac{\delta r\delta z}{3},\\
A_\phi^{11}
&=
r^- z^-
+\frac{r^-\delta z}{2}
+\frac{z^-\delta r}{2}
+\frac{\delta r\delta z}{3}.
\end{aligned}
\end{equation}
In this case, the superscript indices correspond to the radial and axial directions.

For the axial current, the transverse kernels are
\begin{equation}
\begin{aligned}
A_z^{00}
&=
r^+\phi^+
-\frac{r^+\delta\phi}{2}
-\frac{\phi^+\delta r}{2}
+\frac{\delta r\delta\phi}{3},\\
A_z^{10}
&=
r^-\phi^+
-\frac{r^-\delta\phi}{2}
+\frac{\phi^+\delta r}{2}
-\frac{\delta r\delta\phi}{3},\\
A_z^{01}
&=
r^+\phi^-
+\frac{r^+\delta\phi}{2}
-\frac{\phi^-\delta r}{2}
-\frac{\delta r\delta\phi}{3},\\
A_z^{11}
&=
r^-\phi^-
+\frac{r^-\delta\phi}{2}
+\frac{\phi^-\delta r}{2}
+\frac{\delta r\delta\phi}{3}.
\end{aligned}
\end{equation}
The superscript indices now correspond to the radial and azimuthal directions.

The associated swept-volume factors are written as
\begin{equation}
\begin{aligned}
V_{J_r}^{ab}
&=
r_{i+1/2} |\delta r| A_r^{ab},
\qquad a,b\in\{0,1\},\\
V_{J_\phi}^{0b}
&=
r_i |\delta\phi| A_\phi^{0b},
\qquad b\in\{0,1\},\\
V_{J_\phi}^{1b}
&=
r_{i+1} |\delta\phi| A_\phi^{1b},
\qquad b\in\{0,1\},\\
V_{J_z}^{ab}
&=
\widetilde r_p |\delta z| A_z^{ab},
\qquad a,b\in\{0,1\}.
\end{aligned}
\end{equation}
They satisfy the following normalization identities:
\begin{equation}
\begin{aligned}
\sum_{a,b=0}^{1} V_{J_r}^{ab}
&=
r_{i+1/2}\Delta\phi_j\Delta z_k |\delta r|,\\
\sum_{b=0}^{1}V_{J_\phi}^{0b}
+
\sum_{b=0}^{1}V_{J_\phi}^{1b}
&=
\widetilde r_p\Delta r_i\Delta z_k |\delta\phi|,\\
\sum_{a,b=0}^{1} V_{J_z}^{ab}
&=
\widetilde r_p\Delta r_i\Delta\phi_j |\delta z|.
\end{aligned}
\end{equation}

The normalized transverse weights are therefore defined by
\begin{equation}
\begin{aligned}
W_r^{ab}
&=
\frac{A_r^{ab}}{\Delta\phi_j\Delta z_k},
\qquad a,b\in\{0,1\},\\
W_\phi^{ab}
&=
\frac{A_\phi^{ab}}{\Delta r_i\Delta z_k},
\qquad a,b\in\{0,1\},\\
W_z^{ab}
&=
\frac{A_z^{ab}}{\Delta r_i\Delta\phi_j},
\qquad a,b\in\{0,1\}.
\end{aligned}
\label{eq:W}
\end{equation}
For a trajectory segment fully contained in one cell, these weights satisfy
\begin{equation}
\sum_{a,b=0}^{1} W_r^{ab}
=
\sum_{a,b=0}^{1} W_\phi^{ab}
=
\sum_{a,b=0}^{1} W_z^{ab}
=
1 .
\end{equation}

The signed displacement is retained in the current contribution. For the radial current, the contribution of one segment to the face-centered degree of freedom is

\begin{equation}
\begin{aligned}
\Delta J_{r,i+1/2,j+a,k+b}
&=
W_r^{ab}
\frac{q_p w_p \delta r}{\Delta t
r_{i+1/2}\Delta r_i\Delta\phi_j V_z^{k+b}
},\\
\qquad &a,b\in\{0,1\}.
\end{aligned}
\label{eq:delta J r}
\end{equation}

For the azimuthal current, the physical displacement in the azimuthal direction is evaluated at the radial location of the current degree of freedom. Thus, $r_i\delta\phi$ is used at radial node $i$, whereas $r_{i+1}\delta\phi$ is used at radial node $i+1$. The segment contribution is
\begin{equation}
\Delta J_{\phi,i+a,j+1/2,k+b}
=
W_\phi^{ab}
\frac{
q_p w_p r_{i+a}\delta\phi
}{
\Delta t\,
\Delta\phi_j V_r^{i+a} V_z^{k+b}
},
\, a,b\in\{0,1\}.
\end{equation}

The axial-current contribution is
\begin{equation}
\Delta J_{z,i+a,j+b,k+1/2}
=
W_z^{ab}
\frac{
q_p w_p \delta z
}{
\Delta t\,
\Delta\phi_j\Delta z_k V_r^{i+a}
},
\, a,b\in\{0,1\}.
\label{eq:delta J z}
\end{equation}
The total current density at each degree of freedom is obtained by summing these contributions over all particle trajectory segments.

At the cylindrical axis, all azimuthal nodes correspond to the same physical point. The axial current at $r=0$ is therefore regularized by an azimuthally weighted average:
\begin{equation}
J_{z,0,j,k+1/2}
\leftarrow
\frac{
\sum_{\ell} V_\phi^\ell J_{z,0,\ell,k+1/2}
}{
\sum_{\ell} V_\phi^\ell
},
\quad \forall j .
\end{equation}
For a uniform azimuthal mesh, this expression reduces to the arithmetic average over the azimuthal direction.

The resulting deposition scheme preserves the signed charge transport along the particle trajectory while distributing the transverse swept volume according to the local nonuniform cylindrical geometry. The use of cylindrical-volume-weighted measures is essential for maintaining consistency between the deposited current density and the underlying finite-volume discretization on stretched cylindrical grids.

\subsection{Discrete conservation property}
\label{subsec:discrete conservation property}

The conservation property is formulated for the nodal charge before the
node-to-cell projection used in the electrostatic Poisson chain. The proof
therefore concerns the charge-current deposition stage, not the subsequent
cell-centered Poisson source used in the self-field diagnostic. The charge
balance is written on nodal finite-volume control volumes, while the
current components are deposited on staggered degrees of freedom associated
with the corresponding finite-volume fluxes. The derivation below is for
interior off-axis nodes; the periodic azimuthal boundary and the cylindrical
axis are discussed at the end of this subsection.

Let $\Delta t$ be the full particle time step. Consider first one particle
trajectory segment that is fully contained in cell $C_{i,j,k}$. The segment
duration is denoted by $\tau$, with $\tau \leq \Delta t$. For a particle
that does not cross a cell boundary during the time step, $\tau=\Delta t$.
The particle moves from $(r_p,\phi_p,z_p)$ to
$(r'_p,\phi'_p,z'_p)$ along this segment, and the corresponding segment
displacements are
\begin{equation}
\begin{aligned}
\delta r &= r'_p-r_p, \\
\delta\phi &= \phi'_p-\phi_p, \\
\delta z &= z'_p-z_p .
\end{aligned}
\end{equation}
For a surrounding node $(i+a,j+b,k+c)$, where
$a,b,c\in\{0,1\}$, the charge assigned by the first-order shape functions
at a given position on the segment is
\begin{equation}
Q^{(p)}_{abc}(s)
=
Q_p S^r_{i+a}(s)S^\phi_{j+b}(s)S^z_{k+c}(s),
\end{equation}
where $Q_p=q_pw_p$ is the charge represented by the macro-particle.

The change of this nodal charge over the segment is obtained by applying
the product rule along the particle trajectory:
\begin{equation}
\begin{aligned}
\Delta Q^{(p)}_{abc}
&=
Q^{(p)}_{abc}(\tau)-Q^{(p)}_{abc}(0) \\
&=
Q_p
\int_0^{\tau}
\frac{d}{ds}
\left[
S^r_{i+a}(s)S^\phi_{j+b}(s)S^z_{k+c}(s)
\right] ds .
\end{aligned}
\end{equation}
Since the particle trajectory is linear within the segment and the
one-dimensional shape functions are first-order functions, the derivatives
of the shape functions are constant inside the cell:
\begin{equation}
\begin{aligned}
\frac{dS^r_{i+a}}{ds}
&=
(2a-1)\frac{\delta r}{\tau\,\Delta r_i}, \\
\frac{dS^\phi_{j+b}}{ds}
&=
(2b-1)\frac{\delta\phi}{\tau\,\Delta \phi_j}, \\
\frac{dS^z_{k+c}}{ds}
&=
(2c-1)\frac{\delta z}{\tau\,\Delta z_k}.
\end{aligned}
\end{equation}
The transverse time averages appearing in the three product-rule terms are
exactly the weights used in the swept-volume current deposition:
\begin{equation}
\begin{aligned}
W^{bc}_r
&=
\frac{1}{\tau}
\int_0^{\tau}
S^\phi_{j+b}(s)S^z_{k+c}(s)\,ds, \\
W^{ac}_\phi
&=
\frac{1}{\tau}
\int_0^{\tau}
S^r_{i+a}(s)S^z_{k+c}(s)\,ds, \\
W^{ab}_z
&=
\frac{1}{\tau}
\int_0^{\tau}
S^r_{i+a}(s)S^\phi_{j+b}(s)\,ds .
\end{aligned}
\end{equation}
These expressions are equivalent to the normalized swept-volume factors
defined in Eqs.~\ref{eq:Ar}--\ref{eq:W}, with the local distances and displacements
evaluated for the present segment. Therefore, the nodal charge increment
satisfies
\begin{equation}
\begin{aligned}
\frac{\Delta Q^{(p)}_{abc}}{\Delta t}
&=
(2a-1)
\frac{Q_p\delta r}{\Delta t\,\Delta r_i}
W^{bc}_r \\
&\quad+
(2b-1)
\frac{Q_p\delta\phi}{\Delta t\,\Delta\phi_j}
W^{ac}_\phi \\
&\quad+
(2c-1)
\frac{Q_p\delta z}{\Delta t\,\Delta z_k}
W^{ab}_z .
\end{aligned}
\end{equation}

We now define the metric-weighted current fluxes associated with this
particle segment. These fluxes are not independent variables; they are the
finite-volume flux contributions corresponding to the deposited current
densities in Eqs.~\ref{eq:delta J r}--\ref{eq:delta J z}. For the radial current,
\begin{equation}
\begin{aligned}
F^{r,(p)}_{i+1/2,j+b,k+c}
&=
r_{i+1/2}\Delta\phi_j V^z_{k+c}
\Delta J^{(p)}_{r,i+1/2,j+b,k+c} \\
&=
\frac{Q_p\delta r}{\Delta t\,\Delta r_i}
W^{bc}_r .
\end{aligned}
\end{equation}
For the azimuthal current, the corresponding metric-weighted flux is
\begin{equation}
\begin{aligned}
F^{\phi,(p)}_{i+a,j+1/2,k+c}
&=
\frac{V^r_{i+a}V^z_{k+c}}{r_{i+a}}
\Delta J^{(p)}_{\phi,i+a,j+1/2,k+c} \\
&=
\frac{Q_p\delta\phi}{\Delta t\,\Delta\phi_j}
W^{ac}_\phi .
\end{aligned}
\end{equation}
For the axial current,
\begin{equation}
\begin{aligned}
F^{z,(p)}_{i+a,j+b,k+1/2}
&=
V^r_{i+a}\Delta\phi_j
\Delta J^{(p)}_{z,i+a,j+b,k+1/2} \\
&=
\frac{Q_p\delta z}{\Delta t\,\Delta z_k}
W^{ab}_z .
\end{aligned}
\end{equation}

For the surrounding node $(i+a,j+b,k+c)$, the local charge-form discrete
divergence associated with this segment is
\begin{equation}
\begin{aligned}
\mathcal{D}_h J^{(p)}_{abc}
&=
(1-2a)
F^{r,(p)}_{i+1/2,j+b,k+c} \\
&\quad+
(1-2b)
F^{\phi,(p)}_{i+a,j+1/2,k+c} \\
&\quad+
(1-2c)
F^{z,(p)}_{i+a,j+b,k+1/2}.
\end{aligned}
\end{equation}
Substituting the flux expressions above gives
\begin{equation}
\begin{aligned}
\mathcal{D}_h J^{(p)}_{abc}
&=
(1-2a)
\frac{Q_p\delta r}{\Delta t\,\Delta r_i}
W^{bc}_r \\
&\quad+
(1-2b)
\frac{Q_p\delta\phi}{\Delta t\,\Delta\phi_j}
W^{ac}_\phi \\
&\quad+
(1-2c)
\frac{Q_p\delta z}{\Delta t\,\Delta z_k}
W^{ab}_z .
\end{aligned}
\end{equation}
Combining this expression with the nodal charge increment leads to the
local discrete conservation identity
\begin{equation}
\frac{\Delta Q^{(p)}_{abc}}{\Delta t}
+
\mathcal{D}_h J^{(p)}_{abc}
=
0 .
\end{equation}
Thus, for each cell-contained particle segment, the change in the nodal
charge is exactly balanced by the metric-weighted current fluxes deposited
across the corresponding nodal control-volume faces.

For an interior off-axis node, the global charge-form divergence is the
sum of the signed fluxes from the neighboring staggered current degrees of
freedom,
\begin{equation}
\begin{aligned}
\mathcal{D}_h J_{i,j,k}
&=
F^r_{i+1/2,j,k}
-
F^r_{i-1/2,j,k} \\
&\quad+
F^\phi_{i,j+1/2,k}
-
F^\phi_{i,j-1/2,k} \\
&\quad+
F^z_{i,j,k+1/2}
-
F^z_{i,j,k-1/2}.
\end{aligned}
\end{equation}
Here each flux denotes the accumulated charge flux at the corresponding
staggered location. It is obtained by summing the segment-level flux
contributions defined above, using the local metric factors associated
with each segment.

The result extends directly to a full particle time step. If a particle
crosses one or more cell boundaries during $\Delta t$, its trajectory is
split into cell-contained subsegments. For each subsegment, the transverse
weights are evaluated along the local subsegment path, while the current
contribution remains normalized by the full time step $\Delta t$. The
local identity therefore holds for each subsegment in the form
\begin{equation}
\frac{\Delta Q^{(p,m)}_{abc}}{\Delta t}
+
\mathcal{D}_h J^{(p,m)}_{abc}
=
0 .
\end{equation}
When the subsegment contributions are summed, the intermediate charge
assignments at the internal crossing points cancel telescopically, leaving
only the charge difference between the particle position at $t^n$ and that
at $t^{n+1}$. Summing further over all particles gives
\begin{equation}
\frac{Q^{n+1}_{i,j,k}-Q^n_{i,j,k}}{\Delta t}
+
\mathcal{D}_h J_{i,j,k}
=
0 .
\end{equation}
Finally, dividing by the nodal control volume
$V^{\mathrm{node}}_{i,j,k}$ gives the density form of the discrete
continuity equation,
\begin{equation}
\frac{\rho^{n+1}_{i,j,k}-\rho^n_{i,j,k}}{\Delta t}
+
(\nabla_h\cdot \mathbf{J})_{i,j,k}
=
0,
\end{equation}
where
\begin{equation}
\begin{aligned}
\rho_{i,j,k}
&=
\frac{Q_{i,j,k}}{V^{\mathrm{node}}_{i,j,k}}, \\
(\nabla_h\cdot \mathbf{J})_{i,j,k}
&=
\frac{\mathcal{D}_h J_{i,j,k}}
{V^{\mathrm{node}}_{i,j,k}} .
\end{aligned}
\end{equation}

In the periodic azimuthal direction, the particle trajectory is locally
unwrapped before the weights are evaluated, and the deposited quantities
are then accumulated with periodic indexing. The flux crossing the
periodic boundary is therefore treated as an internal flux and does not
introduce an additional source term. If the cylindrical axis is included,
the axis is treated as a geometrically degenerate point. The conservation
statement is then applied to the azimuthally combined axis charge,
consistent with the axis charge accumulation and axial-current
regularization used in the deposition scheme.

\subsection{Implementation Layout for the Poisson--Gradient--Gather Chains}

The PIC tests are performed on a logically structured three-dimensional
cylindrical mesh with coordinates $(r,\phi,z)$, where $\phi$ denotes the
azimuthal angle. The mesh is allowed to be nonuniform in the radial,
azimuthal, and axial directions.

Sect.\ref{subsec:geometric_measures_nonuniform_cylindrical}--\ref{subsec:discrete conservation property} describe the geometric measures used in the deposition
derivation, where $r_i$, $\phi_j$, and $z_k$ denote the nodal partition
associated with nodal charge control volumes. In this subsection, we switch to
the indexing convention used in the implementation of the electrostatic
Poisson--gradient--gather chain. Cell faces are denoted by
$r^{\rm f}_{i-1/2}$, $\phi^{\rm f}_{j-1/2}$, and
$z^{\rm f}_{k-1/2}$, while cell centers are denoted by
$r^{\rm c}_i$, $\phi^{\rm c}_j$, and $z^{\rm c}_k$. This implementation
notation is used only to describe the cell-centered Poisson source, the
cell-centered potential, the face-centered electric field, and the field gather
operation. It does not redefine the nodal control volumes introduced in
Sec.~\ref{subsec:geometric_measures_nonuniform_cylindrical}.

The corresponding local cell widths are
\begin{equation}
\begin{array}{rcl}
\Delta r_i &=& r^{\rm f}_{i+1/2}-r^{\rm f}_{i-1/2},  \\[3pt]
\Delta \phi_j &=& \phi^{\rm f}_{j+1/2}-\phi^{\rm f}_{j-1/2}, \\[3pt]
\Delta z_k &=& z^{\rm f}_{k+1/2}-z^{\rm f}_{k-1/2}.
\end{array}
\end{equation}

The azimuthal coordinate is stored as an angle. If the input azimuthal span is
specified as an arc length, it is converted to an angular width using a
reference radius $r_c=(r_{\min}+r_{\max})/2$. The azimuthal direction is
treated as periodic unless otherwise stated.

Scalar quantities are stored at cell centers. These quantities include the
electrostatic potential $\Phi$, charge density $\rho$, electron number
density $n_e$, and ion number density $n_i$. The cell-center coordinates are
defined as
\begin{equation}
\begin{array}{rcl}
r^{\rm c}_i &=&
\dfrac{r^{\rm f}_{i-1/2}+r^{\rm f}_{i+1/2}}{2}, \\[6pt]
\phi^{\rm c}_j &=&
\dfrac{\phi^{\rm f}_{j-1/2}+\phi^{\rm f}_{j+1/2}}{2}, \\[6pt]
z^{\rm c}_k &=&
\dfrac{z^{\rm f}_{k-1/2}+z^{\rm f}_{k+1/2}}{2}.
\end{array}
\end{equation}

The finite-volume cell volume used in the Poisson equation is evaluated with
the cylindrical Jacobian included. For cell $(i,j,k)$, the exact volume is
\begin{equation}
V_{i,j,k}
=
\frac{1}{2}
\left[
\left(r^{\rm f}_{i+1/2}\right)^2
-
\left(r^{\rm f}_{i-1/2}\right)^2
\right]
\Delta \phi_j \Delta z_k .
\end{equation}
With the arithmetic radial cell center defined above, this volume can also be
written as
\begin{equation}
V_{i,j,k}
=
r^{\rm c}_i\,\Delta r_i\,\Delta \phi_j\,\Delta z_k .
\end{equation}
Thus, the midpoint-radius form used in the implementation is not an additional
approximation; it is equivalent to the exact cylindrical finite-volume measure
under the present definition of $r^{\rm c}_i$.

The electrostatic Poisson equation is discretized in a cell-centered
finite-volume form, and the electrostatic potential is solved at cell
centers. This layout assigns one scalar unknown to each control volume,
allows the projected cell-centered charge density to be used directly as
the Poisson source term, and imposes Dirichlet data through boundary
faces rather than shared corner nodes. It therefore avoids the need for
an additional corner-node convention when adjacent boundary segments
carry different prescribed potentials.
The electric field is then placed on a staggered grid, with each
electric-field component computed on the face normal to its own
direction. This staggered placement also facilitates the finite-volume
gradient evaluation, because the potential difference between two
neighboring cell centers directly gives the normal electric field on
their common face. The radial electric field is located at
$(i+1/2,j,k)$, the azimuthal electric field at $(i,j+1/2,k)$, and the
axial electric field at $(i,j,k+1/2)$. After the cell-centered potential
is obtained, the face-centered electric fields are evaluated as
\begin{equation}
E_r\big|_{i+1/2,j,k}
=
-
\frac{\Phi_{i+1,j,k}-\Phi_{i,j,k}}
{r^{\rm c}_{i+1}-r^{\rm c}_{i}}
=
-
\frac{\Phi_{i+1,j,k}-\Phi_{i,j,k}}
{\frac{1}{2}(\Delta r_i+\Delta r_{i+1})},
\end{equation}
\begin{equation}
E_\phi\big|_{i,j+1/2,k}
=
-
\frac{1}{r^{\rm c}_i}
\frac{\Phi_{i,j+1,k}-\Phi_{i,j,k}}
{\phi^{\rm c}_{j+1}-\phi^{\rm c}_{j}}
=
-
\frac{1}{r^{\rm c}_i}
\frac{\Phi_{i,j+1,k}-\Phi_{i,j,k}}
{\frac{1}{2}(\Delta \phi_j+\Delta \phi_{j+1})},
\end{equation}
and
\begin{equation}
E_z\big|_{i,j,k+1/2}
=
-
\frac{\Phi_{i,j,k+1}-\Phi_{i,j,k}}
{z^{\rm c}_{k+1}-z^{\rm c}_{k}}
=
-
\frac{\Phi_{i,j,k+1}-\Phi_{i,j,k}}
{\frac{1}{2}(\Delta z_k+\Delta z_{k+1})}.
\end{equation}
For diagnostics and visualization, cell-centered electric fields are obtained
by averaging the two adjacent face-centered values in each direction.

Particle charge is first deposited to the surrounding eight nodal points using
first-order trilinear shape functions based on the local nonuniform cell
widths. These nodal points and their control volumes follow the geometric
definition in Sec.~\ref{subsec:geometric_measures_nonuniform_cylindrical}. The nodal charge contribution is normalized by the
corresponding nodal control volume. 
After all particle contributions have been accumulated at the nodal
points, the nodal density is projected to cell centers to obtain
$n_e$, $n_i$, and $\rho=q_en_e+q_in_i$.
This node-to-cell transfer provides the
cell-centered source term required by the Poisson solver, while the
charge-current conservation property discussed in Sec.~\ref{subsec:discrete conservation property} is formulated at
the nodal charge level before this projection.
The node-to-cell transfer used in the Poisson solve is a
cylindrical-volume-weighted projection rather than an arithmetic
eight-node average. For a cell \(C_{i,j,k}\), let
\(r_L\) and \(r_R\) denote its left and right radial faces. The
cell-centered charge density is obtained as
\begin{equation}
\begin{aligned}
\rho^{\rm cc}_{i,j,k}
=&\frac{w_L}{4}
\sum_{b=0}^{1}\sum_{c=0}^{1}
\rho^{\rm node}_{i,j+b,k+c}  \\
&+\frac{w_R}{4}
\sum_{b=0}^{1}\sum_{c=0}^{1}
\rho^{\rm node}_{i+1,j+b,k+c} ,
\end{aligned}
\end{equation}
where the radial weights are defined by the volume average of the
linear radial basis functions,
\begin{equation}
\begin{aligned}
w_L &=
\frac{\int_{r_L}^{r_R} r N_L(r)\,dr}
     {\int_{r_L}^{r_R} r\,dr}
=
\frac{2r_L+r_R}{3(r_L+r_R)},\\
w_R &=
\frac{\int_{r_L}^{r_R} r N_R(r)\,dr}
     {\int_{r_L}^{r_R} r\,dr}
=
\frac{r_L+2r_R}{3(r_L+r_R)} .
\end{aligned}
\end{equation}
Here the factors \(1/4\) come from the ordinary averages of the
linear basis functions in the \(\phi\)- and \(z\)-directions. Since
\(w_L+w_R=1\), the projection preserves constant density fields.
When the same nodal basis and nodal control volumes are used, it also
preserves the total charge in the finite-volume sense,
\begin{equation}
\sum_{i,j,k} \Omega_{i,j,k}\rho^{\rm cc}_{i,j,k}
=
\sum_{i,j,k} V^{\rm node}_{i,j,k}\rho^{\rm node}_{i,j,k}.
\end{equation}
In the implementation, \(n_e\) and \(n_i\) are projected separately by
this linear operator, and the cell-centered charge density is then
formed as
\begin{equation}
\rho^{\rm cc}=q_e n_e^{\rm cc}+q_i n_i^{\rm cc}.
\end{equation}

During particle pushing, the face-centered electric field is interpolated back
to the particle position. Each component is interpolated from its own staggered
location: $E_r$ from radial faces, $E_\phi$ from azimuthal faces, and
$E_z$ from axial faces. The magnetic field used in the Boris pusher is
defined on an $r$-$z$ nodal grid and interpolated to the particle position
when required. This staggered arrangement reduces ambiguity in field
interpolation and keeps the field construction consistent with the
finite-volume discretization of the cylindrical Poisson equation.

\section{Verification of Charge and Current Density Deposition}

\subsection{Verification of Charge Deposition}

The charge deposition scheme was first verified by examining whether a
uniform particle distribution in physical volume can be recovered as a uniform
charge density on cylindrical grids. The computational domain was a unit
cylinder,
\begin{equation}
0 \leq r \leq 1,\quad
0 \leq \phi < 2\pi,\quad
0 \leq z \leq 1 .
\end{equation}
A total of $N_p=2\times 10^8$ particles were used, with $q_p=1$ and
$w_p=1$. To obtain a uniform distribution in cylindrical volume, the particle
positions were sampled as
\begin{equation}
r=\sqrt{\xi_r},\quad
\phi=2\pi\xi_\phi,\quad
z=\xi_z ,
\end{equation}
where $\xi_r$, $\xi_\phi$, and $\xi_z$ are independent random numbers
uniformly distributed in $[0,1)$. Since the volume of the unit cylinder is
$\pi$, the corresponding reference charge density is
\begin{equation}
\rho_0=\frac{N_p q_p w_p}{\pi}
      \simeq 6.37\times 10^7 .
\end{equation}

The particle charge was deposited to the surrounding eight nodes using the
first-order shape functions described above. The nodal charge was normalized by
the corresponding nodal control volume to obtain the charge density. Since all
azimuthal nodes at the cylindrical axis represent the same physical location,
the deposited axis charge was averaged in the azimuthal direction before the
axis density was evaluated.

\begin{figure}
    \centering
    \includegraphics[width=0.7\linewidth]{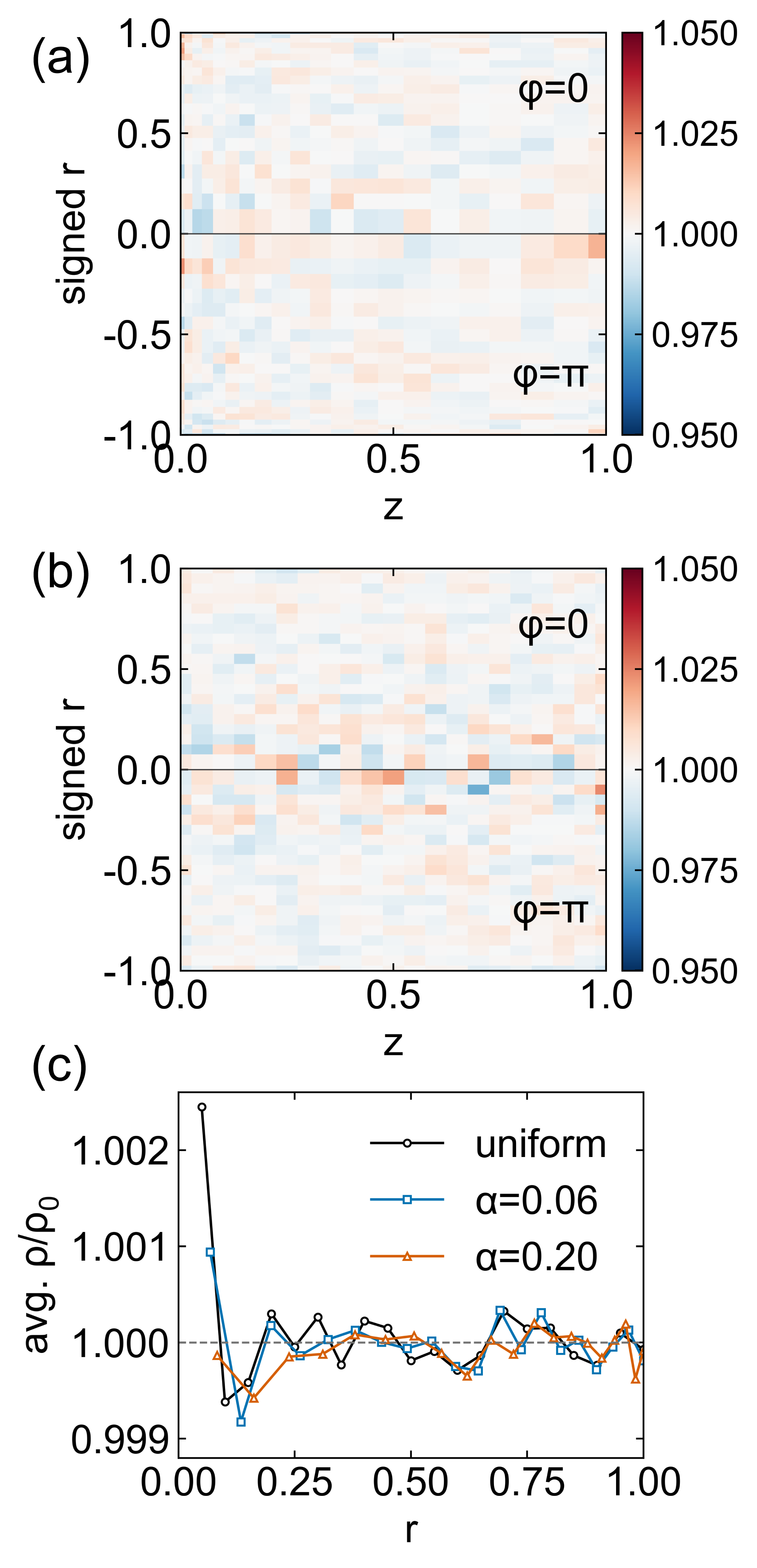}
    \caption{Charge-density deposition verification on cylindrical grids.
(a) and (b) show the normalized density $\rho/\rho_0$ on
representative off-axis $r$-$z$ slices for the uniform and strongly
stretched grids, respectively. In these slice plots, a signed radial coordinate is used only for visualization: the upper half corresponds to the $\phi=0$ plane and the lower half to the $\phi=\pi$ plane, while the physical radius remains nonnegative. (c) shows the azimuthally and axially
averaged radial profiles for $\alpha=0$, $0.06$, and $0.20$.}
    \label{fig:charge deposition}
\end{figure}

For the uniform-grid case, the mesh resolution was
$N_r=20$, $N_\phi=20$, and $N_z=20$, corresponding to
$\Delta r=0.05$, $\Delta\phi=\pi/10$, and $\Delta z=0.05$.
The deposited density was compared with $\rho_0$ on representative
$r$-$z$ planes at $\phi=0$ and $\phi=\pi$, and the radial density
profile was further averaged over the azimuthal and axial directions. Because
the cylindrical axis is a geometrically degenerate point, the quantitative
comparison below excludes the axis node. On the uniform grid, the off-axis
$r$-$z$ slice varies by only a few percent around $\rho_0$, and the
azimuthally and axially averaged radial profile is nearly flat.

The same test was then repeated on nonuniform cylindrical meshes. The grid
was refined away from the cylindrical axis in the radial direction,
\begin{equation}
\begin{aligned}
\Delta r_i &= \Delta r_0[1+\alpha(N_r-1-i)],\\
\Delta \phi_j &= \Delta \phi_0(1+\alpha j),\\
\Delta z_k &= \Delta z_0(1+\alpha k),
\end{aligned}
\end{equation}
where $\alpha$ controls the stretching strength. Three cases were considered:
$\alpha=0$, $\alpha=0.06$, and $\alpha=0.20$. The case $\alpha=0$
corresponds to the uniform-grid reference obtained with the same nonuniform-grid
formulation.

The corresponding density slices and radial profiles are shown in Fig.~\ref{fig:charge deposition}.
For the strongly stretched case with $\alpha=0.20$, the representative
off-axis $r$-$z$ slice gives
$0.986\leq\rho/\rho_0\leq1.023$. The radial profiles for all three grids stay
within $0.999$--$1.002$, as summarized in
Tab.~\ref{tab:charge_deposition_verification}. Thus, reversing the radial
stretching so that the outer region is more finely resolved does not introduce a
systematic bias in the off-axis charge deposition.

\begin{table}[t]
\centering
\caption{Off-axis charge-density ranges for uniform and far-from-axis-refined cylindrical grids.}
\begin{tabular}{lccc}
\hline
Grid & Radial spacing & $r$-$z$ slice & Radial profile \\
& & $\rho/\rho_0$ & $\langle\rho\rangle_{\phi,z}/\rho_0$ \\
\hline
Uniform, $\alpha=0$ & uniform & 0.975--1.024 & 0.9994--1.0024 \\
Nonuniform, $\alpha=0.06$ & outer refined & 0.985--1.019 & 0.9992--1.0009 \\
Nonuniform, $\alpha=0.20$ & outer refined & 0.986--1.023 & 0.9994--1.0002 \\
\hline
\end{tabular}
\label{tab:charge_deposition_verification}
\end{table}

\subsection{Verification of Current Deposition}

The current deposition scheme was verified using a controlled transport test in
the same cylindrical domain as the charge deposition test. Particles were
initially distributed uniformly in physical volume inside
\begin{equation}
0 \leq r \leq 1,\qquad
0 \leq \phi < 2\pi,\qquad
0 \leq z \leq 1 .
\end{equation}
The initial positions were sampled by $r=\sqrt{\xi_r}$,
$\phi=2\pi\xi_\phi$, and $z=\xi_z$, so that the reference charge density
was spatially uniform, with $\rho_0=N_pq_pw_p/\pi$.

\begin{figure}
    \centering
    \includegraphics[width=0.7\linewidth]{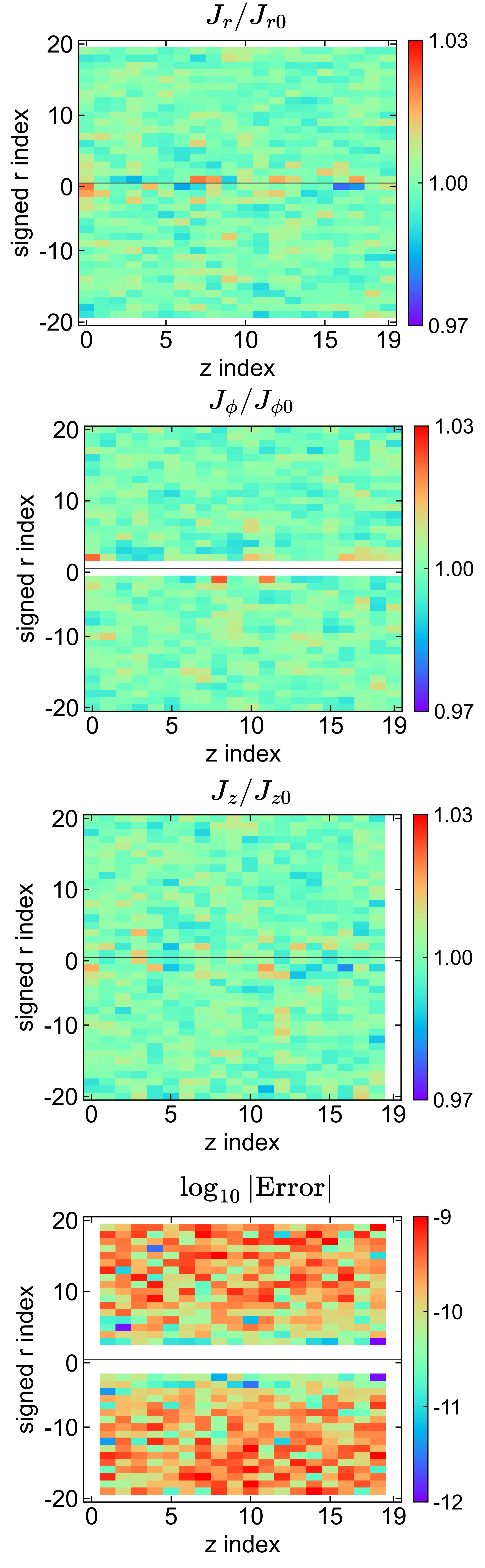}
    \caption{Current-density deposition verification on the strongly stretched cylindrical grid with $\alpha=0.20$. From top to bottom, the panels show the normalized deposited current components $J_r/J_r^{\rm ref}$, $J_\phi/J_\phi^{\rm ref}$, and $J_z/J_z^{\rm ref}$, followed by the base-10 logarithm of the absolute discrete-continuity residual.}
    \label{fig:current deposition}
\end{figure}

Each particle was then transported by a prescribed small displacement during
one time step. In the uniform-grid test, the imposed displacement was
$\delta r=5.0\times10^{-5}$, $\delta\phi=3.1416\times10^{-4}$, and
$\delta z=5.0\times10^{-5}$, with $\Delta t=0.01$. The corresponding
radial velocity, angular velocity, and axial velocity were
$u_r=5.0\times10^{-3}$, $\dot{\phi}=3.1416\times10^{-2}$, and
$u_z=5.0\times10^{-3}$, respectively. For a uniform reference density, the
expected current density is
\begin{equation}
J_r^{\mathrm{ref}}=\rho_0 u_r,\quad
J_\phi^{\mathrm{ref}}=\rho_0 r\dot{\phi},\quad
J_z^{\mathrm{ref}}=\rho_0 u_z .
\end{equation}

The deposited current was assessed in two ways. First, each component of the
current density was compared with its analytical reference value. Second, the
deposited current was substituted into the discrete continuity equation,
\begin{equation}
\frac{\rho^{n+1}-\rho^n}{\Delta t}
+
\nabla_h\cdot\mathbf{J}
=0,
\end{equation}
where the discrete divergence operator used the same cylindrical
finite-volume metric factors as the deposition scheme. This provides a direct
test of whether the current deposition is consistent with the corresponding
charge update.

For the uniform-grid case, $N_p=2\times10^8$ particles were used on a
$20\times20\times20$ mesh. The normalized RMS errors of the three current
components are all below $7\times10^{-3}$. The maximum pointwise errors are
larger because they retain local particle-count fluctuations, but the
continuity residual remains much smaller than the current-density error. This
indicates that the deposited current is locally consistent with the charge
variation over one time step.

The same test was then repeated on radially nonuniform cylindrical meshes. The
azimuthal and axial spacings were kept uniform, while the radial spacing was
reversed so that the outer radial region is more finely resolved,
\begin{equation}
\Delta r_i=\Delta r_0[1+\alpha(N_r-1-i)],
\end{equation}
where $\alpha$ controls the radial stretching strength. Two stretched cases,
$\alpha=0.06$ and $\alpha=0.20$, were considered to isolate the influence
of the nonuniform radial metric.

Fig.~\ref{fig:current deposition} shows the normalized current components and the discrete-continuity residual for the strongly stretched case.
For $\alpha=0.20$, the representative $r$-$z$ slices give
$0.965\leq J_r/J_r^{\mathrm{ref}}\leq1.029$,
$0.977\leq J_\phi/J_\phi^{\mathrm{ref}}\leq1.023$, and
$0.974\leq J_z/J_z^{\mathrm{ref}}\leq1.018$. The RMS errors of all three
current components remain at the level of $10^{-3}$ for the uniform and
nonuniform grids, as shown in Tab.~\ref{tab:current_deposition_verification}.
The small differences among the three grids are comparable to the sampling
scatter of the particle realization. The important point is that the
continuity residual stays several orders of magnitude below the current-density
error, showing that the current deposition preserves the discrete charge
balance on the far-from-axis-refined mesh.

\begin{table}[t]
\centering
\caption{Normalized RMS errors of current deposition on uniform and outer-refined cylindrical grids.}
\begin{tabular}{lccc}
\hline
Error metric
& Uniform
& outer $\alpha=0.06$
& outer $\alpha=0.20$ \\
\hline
RMS $J_r$ error
& $6.43\times10^{-3}$
& $5.55\times10^{-3}$
& $4.98\times10^{-3}$ \\

RMS $J_\phi$ error
& $5.79\times10^{-3}$
& $4.86\times10^{-3}$
& $4.62\times10^{-3}$ \\

RMS $J_z$ error
& $5.74\times10^{-3}$
& $4.86\times10^{-3}$
& $4.50\times10^{-3}$ \\

Max. cont. residual
& $1.44\times10^{-9}$
& $9.19\times10^{-10}$
& $2.25\times10^{-9}$ \\
\hline
\end{tabular}
\label{tab:current_deposition_verification}
\end{table}

\section{Numerical Self-Field Characterization}

\subsection{Effect of Field Placement: Face-Centered, Cell-Centered, and Shifted Cell-Centered Layouts}

\begin{figure}
    \centering
    \includegraphics[width=0.8\linewidth]{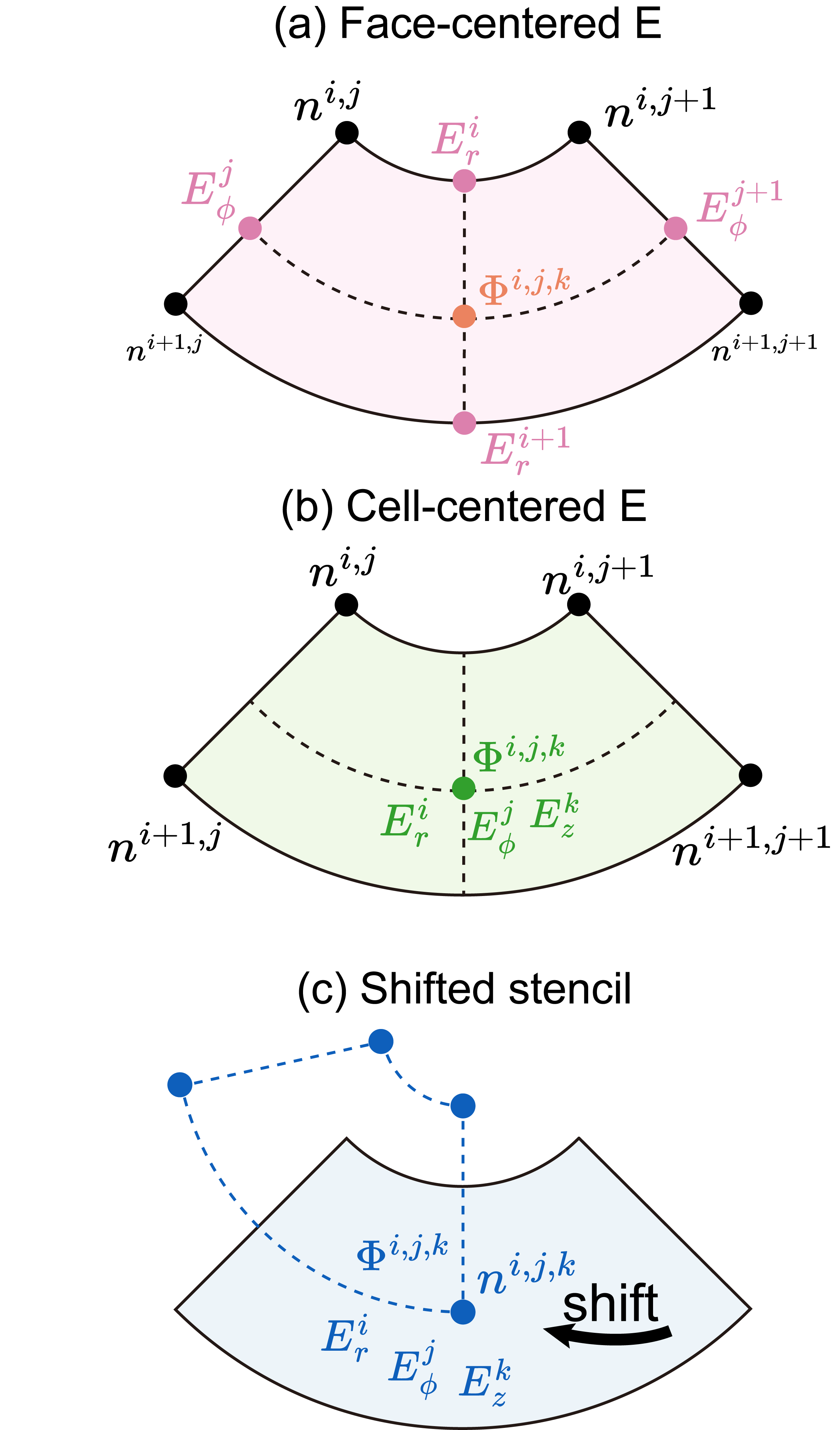}
    \caption{Field-placement layouts for the self-field test.}
    \label{fig:field placement}
\end{figure}

The self-field diagnostics in this section were performed using
AlgoPlasma, a plasma-simulation library developed by the
authors and their research team (previously referred to as PMSL\cite{Chen2025PoP_PMSLPICHET3D,Zhao2025IEPC_MagneticFieldConfiguration,chen2025iepcMCCNeutralSolver,zhong2026effects}) and
prepared for a future open-source release. The name AlgoPlasma reflects
the design goal of the library: to make the numerical algorithms in
plasma simulation explicit, modular, and testable. In particular,
deposition, Poisson solve, field reconstruction, gather interpolation,
and particle pushing are implemented as separable algorithmic
components, allowing different particle--field coupling chains to be
systematically examined.

To assess the influence of field placement on the residual numerical self-field, three layouts are compared on the same nonuniform cylindrical mesh. 
The three field-placement layouts compared here are sketched in Fig.~\ref{fig:field placement}.
The comparison is designed to isolate the effect of the consistency among charge deposition, Poisson discretization, field reconstruction, and particle gather.

The computational domain is a three-dimensional cylindrical domain with $N_r \times N_\phi \times N_z = 16 \times 16 \times 16$. The physical ranges are $r \in [2.0\times 10^{-2}, 5.0\times 10^{-2}]~\mathrm{m}$, $\phi \in [0, 5.71\times 10^{-1}]~\mathrm{rad}$, and $z \in [0, 3.0\times 10^{-2}]~\mathrm{m}$. The azimuthal extent corresponds to an arc length of $0.02~\mathrm{m}$ at $r_{\mathrm{center}} = 0.035~\mathrm{m}$.

A nonuniform mesh is employed in all three directions. The radial grid is generated with power $1.8$ and lower-side focusing, the azimuthal grid with power $1.4$ and both-side focusing, and the axial grid with power $1.6$ and upper-side focusing. Zero Dirichlet boundary conditions are imposed on all six boundaries.

The test case contains a single $\mathrm{Xe}^{+}$ ion with particle weight $w_i = 1$, without electrons. The sampling cell is $(i,j,k) = (8,8,8)$. The self-field is evaluated on three orthogonal planes passing through the equal-deposition point, namely the $r-\phi$, $r-z$, and $\phi-z$ planes. Each plane contains $100 \times 100$ particle positions, yielding $30000$ sampling points in total.

The equal-deposition point is located at
$(\xi_r,\xi_\phi,\xi_z)=(0.543,0.507,0.482)$ in logical coordinates,
corresponding to
$(r,\phi,z)=(2.78\times 10^{-2}~\mathrm{m},2.62\times 10^{-1}~\mathrm{rad},
1.90\times 10^{-2}~\mathrm{m})$ in physical coordinates.

The residual self-field is measured by the gathered field magnitude
$|E_{\mathrm{self}}|$. To estimate its direct effect on particle motion, a
dimensionless one-step displacement is also used:
\begin{equation}
K_x =
\frac{1}{2}
\frac{q_i |E_{\mathrm{self}}|}{m_i}
\frac{\Delta t^2}{h_{\mathrm{eff}}},
\end{equation}
where $h_{\mathrm{eff}}\simeq1.72\times10^{-3}~\mathrm{m}$ is the local
effective mesh size and $\Delta t=10^{-12}~\mathrm{s}$.

The first layout is the baseline face-centered formulation used in the main solver:
\begin{equation}
\mathrm{particle}
\rightarrow
\rho_{\mathrm{node}}
\rightarrow
\rho_{\mathrm{cc}}
\rightarrow
\Phi_{\mathrm{cc}}
\rightarrow
E_{\mathrm{face}}
\rightarrow
E_{\mathrm{particle}} .
\end{equation}
The second layout retains the same deposition and Poisson solve, but reconstructs the electric field at cell centers:
\begin{equation}
\mathrm{particle}
\rightarrow
\rho_{\mathrm{node}}
\rightarrow
\rho_{\mathrm{cc}}
\rightarrow
\Phi_{\mathrm{cc}}
\rightarrow
E_{\mathrm{cc}}
\rightarrow
E_{\mathrm{particle}} .
\end{equation}
The third layout further shifts the deposition to a cell-centered representation:
\begin{equation}
\mathrm{particle}
\rightarrow
\rho_{\mathrm{cc}}
\rightarrow
\Phi_{\mathrm{cc}}
\rightarrow
E_{\mathrm{cc}}
\rightarrow
E_{\mathrm{particle}} .
\end{equation}

The statistical results over all $30000$ sampling points are summarized in
Tab.~\ref{tab:self_field_layout_statistics}. Among the three layouts, the
face-centered formulation gives the smallest residual self-field, with
$|E_{\mathrm{self}}|_{\mathrm{RMS}}=1.02\times10^{-5}~\mathrm{V/m}$. The
cell-centered reconstruction increases this value by a factor of about $2.2$,
and the shifted cell-centered layout increases it by a factor of about $4.4$.
The same ordering is obtained from $K_x$.

\begin{table}[htbp]
\centering
\footnotesize
\caption{Residual self-field statistics for different field layouts.}
\label{tab:self_field_layout_statistics}
\setlength{\tabcolsep}{3.0pt}
\begin{tabular}{lccccc}
\hline
Layout &
$h_{\mathrm{eff}}$ &
$E_{\mathrm{RMS}}$ &
$E_{\max}$ &
$K_{x,\mathrm{RMS}}$ &
$K_{x,\max}$ \\
&
$(10^{-3}~\mathrm{m})$ &
\multicolumn{2}{c}{$(10^{-5}~\mathrm{V/m})$} &
\multicolumn{2}{c}{$(10^{-21})$} \\
\hline
Face-centered $E$ & 1.72 & 1.02 & 2.19 & 2.17 & 4.67 \\
Cell-centered $E$ & 1.72 & 2.27 & 3.65 & 4.84 & 7.80 \\
Shifted cell-centered & 1.72 & 4.51 & 9.64 & 9.63 & 20.59 \\
\hline
\end{tabular}
\end{table}

\begin{figure}
    \centering
    \includegraphics[width=\linewidth]{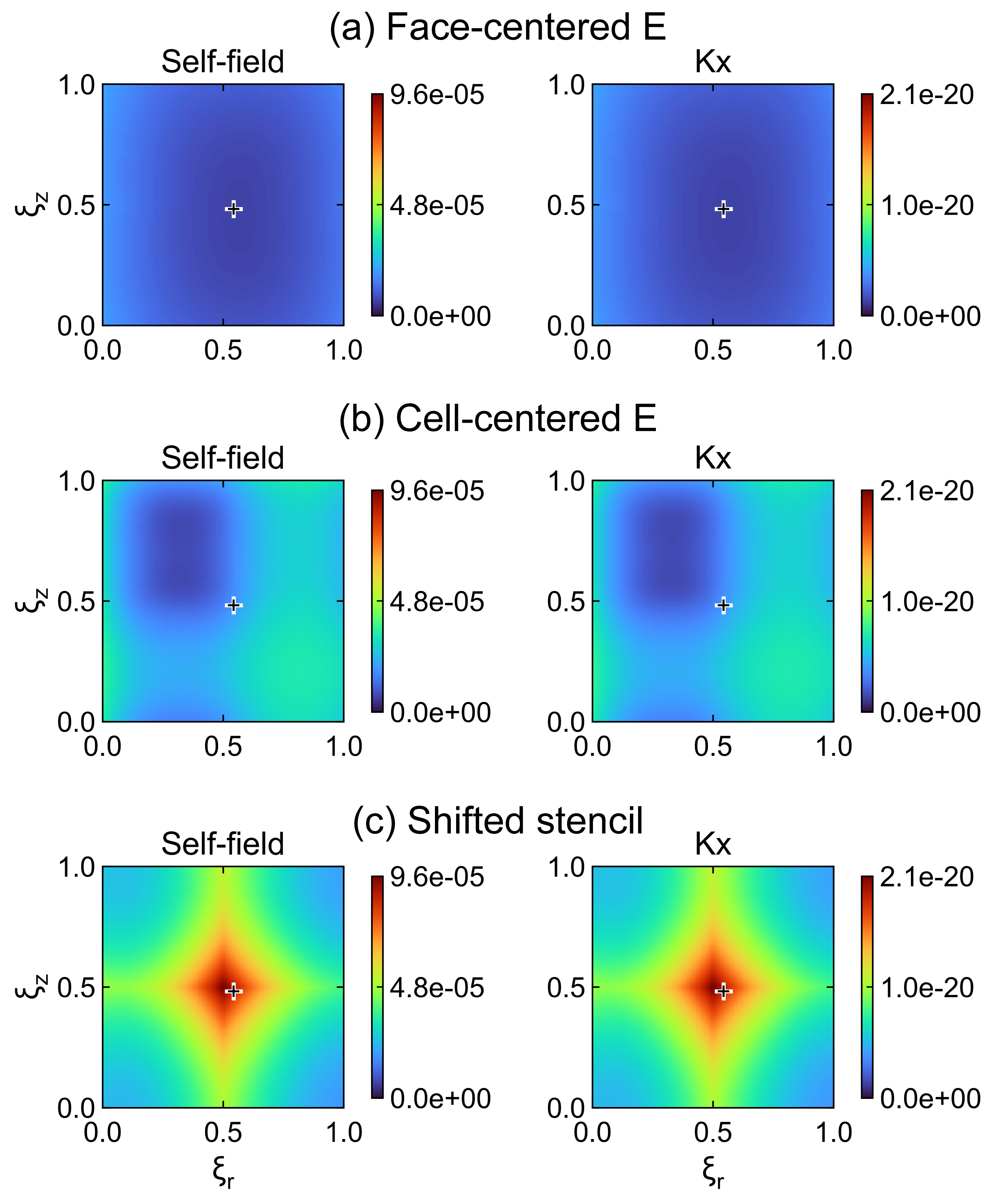}
    \caption{Residual self-field and normalized one-step displacement $K_x$ for different electric-field placement layouts on the same nonuniform cylindrical mesh. Rows (a)--(c) correspond to the face-centered, cell-centered, and shifted cell-centered layouts, respectively. The plotted slice is the $r$-$z$ sampling plane passing through the equal-deposition point.}
    \label{fig:self force}
\end{figure}


The results indicate that the residual self-field is not controlled solely by the collocation of $\Phi$ and $E$. Instead, it is more strongly affected by the compatibility among the deposition scheme, the Poisson discretization, the discrete gradient operator, and the gather procedure. In the present nonuniform cylindrical formulation, the face-centered electric field is obtained from $\Phi_{\mathrm{cc}}$ through the natural finite-volume discrete gradient, and this pairing gives the smallest residual self-field.

By contrast, reconstructing $E$ at cell centers from $\Phi_{\mathrm{cc}}$ does not reduce the self-field; instead, it increases it substantially. A further shift of the deposition to a cell-centered representation leads to an even larger residual self-field. Therefore, for the present nonuniform cylindrical mesh, the face-centered layout is the most favorable among the three candidates and is adopted as the baseline in the following analysis.

The $r$-$z$ slices in Fig.~\ref{fig:self force} show the same trend in
space: the face-centered layout exhibits the weakest residual self-field, while
the shifted cell-centered layout exhibits the strongest. The values of $K_x$
remain far below unity in all three cases, so the one-step displacement caused
by the residual self-field is much smaller than the local mesh size.
Because all three field-placement layouts are compared under the same domain size, particle location, and boundary condition, the relative differences reported here mainly reflect the effect of field placement within this fixed electrostatic chain. Boundary-induced contributions may affect the absolute self-field magnitude, but they do not change the purpose of the present comparison.


\subsection{Effect of Grid Nonuniformity in the Cartesian-Limit Cylindrical Geometry}

The previous subsection indicates that the face-centered electric-field layout gives the lowest residual self-field among the tested field-placement schemes. This subsection further separates the effect of cylindrical curvature from that of grid nonuniformity. The radial coordinate is shifted to a much larger radius, with $r_{\min}=200~\mathrm{m}$, so that the local cylindrical geometry approaches the Cartesian limit.

All cases use the same baseline finite-volume sequence,
\begin{equation}
\mathrm{particle}
\rightarrow
\rho_{\mathrm{node}}
\rightarrow
\rho_{\mathrm{cc}}
\rightarrow
\Phi_{\mathrm{cc}}
\rightarrow
E_{\mathrm{face}}
\rightarrow
E_{\mathrm{particle}} .
\end{equation}
The transfer from $\rho_{\mathrm{node}}$ to $\rho_{\mathrm{cc}}$ is performed using the corrected cylindrical-volume-weighted node-to-cell formulation.

\begin{figure}
    \centering
    \includegraphics[width=\linewidth]{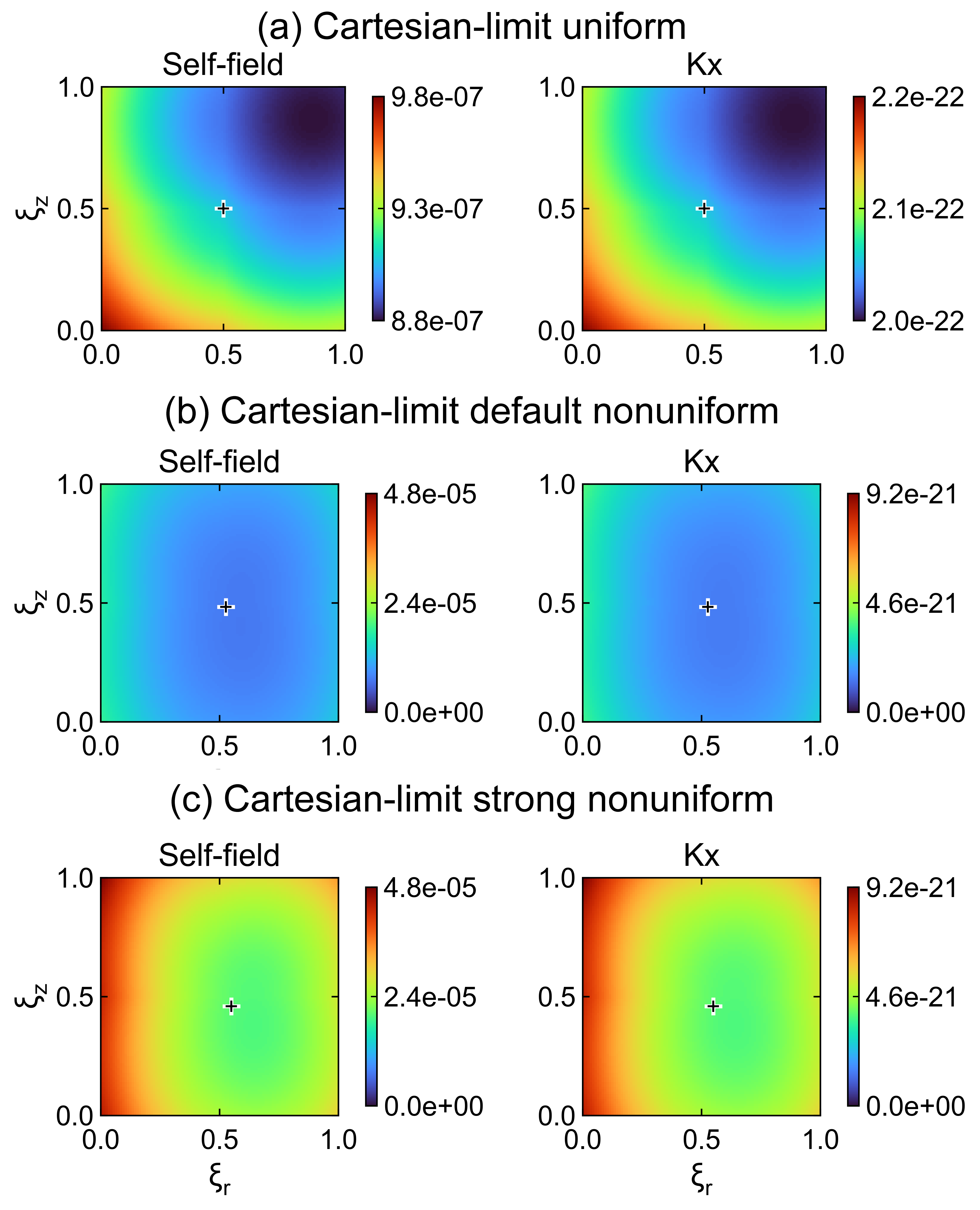}
    \caption{Residual self-field and normalized one-step displacement $K_x$ in the Cartesian-limit cylindrical geometry. Rows (a)--(c) correspond to the uniform, default nonuniform, and strongly nonuniform grids, respectively. The plotted slice is the $r$-$z$ sampling plane passing through the equal-deposition point.}
    \label{fig:Cartesian-limit self force}
\end{figure}

The computational domain is still described in three-dimensional cylindrical coordinates, with $N_r \times N_\phi \times N_z = 16 \times 16 \times 16$. The radial and axial ranges are $r \in [200.0,200.03]~\mathrm{m}$ and $z \in [0,0.03]~\mathrm{m}$, respectively. The azimuthal range is determined from an arc length of $0.02~\mathrm{m}$ at the center radius:
\begin{equation}
\phi_{\max}
=
\frac{0.02}{r_{\mathrm{center}}}
\simeq
1.00\times 10^{-4}~\mathrm{rad}.
\end{equation}
Zero Dirichlet boundary conditions are imposed on all six boundaries.

The test particle is a single $\mathrm{Xe}^{+}$ ion with $w_i=1$, and no electrons are included. The sampling cell is $(i,j,k)=(8,8,8)$. The residual self-field is evaluated on the $r-\phi$, $r-z$, and $\phi-z$ planes passing through the equal-deposition point. Each plane contains $100\times 100$ particle positions, giving $30000$ sampling points in total.

Three levels of grid nonuniformity are considered. The first case uses a uniform grid in all three directions. The second case uses the default nonuniform grid, with power $1.8$ and lower-side focusing in $r$, power $1.4$ and both-side focusing in $\phi$, and power $1.6$ and upper-side focusing in $z$. The third case uses a stronger nonuniform grid, with power $2.5$ and lower-side focusing in $r$, power $2.2$ and both-side focusing in $\phi$, and power $2.4$ and upper-side focusing in $z$.

The self-field statistics are shown in Fig.~\ref{fig:Cartesian-limit self force}
and summarized in
Tab.~\ref{tab:cartesian_limit_statistics}. In the nearly Cartesian geometry,
the uniform grid gives a small residual self-field, with
$|E_{\mathrm{self}}|_{\mathrm{RMS}}=1.00\times10^{-6}~\mathrm{V/m}$. The
default nonuniform grid increases the RMS self-field by about one order of
magnitude, and the strong nonuniform grid increases it by about a factor of
$26$. The normalized displacement $K_x$ follows the same ordering.

\begin{table}[htbp]
\centering
\footnotesize
\caption{Self-field statistics in the Cartesian-limit cylindrical geometry.}
\label{tab:cartesian_limit_statistics}
\setlength{\tabcolsep}{2.0pt}
\begin{tabular}{lccccc}
\hline
Case &
$h_{\mathrm{eff}}$ &
$E_{\mathrm{RMS}}$ &
$E_{\max}$ &
$K_{x,\mathrm{RMS}}$ &
$K_{x,\max}$ \tabularnewline
&
$(10^{-3}~\mathrm{m})$ &
\multicolumn{2}{c}{$(10^{-6}~\mathrm{V/m})$} &
\multicolumn{2}{c}{$(10^{-21})$} \tabularnewline
\hline
Uniform & 1.64 & 1.00 & 1.79 & 0.225 & 0.402 \tabularnewline
Default & 1.86 & 9.93 & 19.20 & 1.96 & 3.79 \tabularnewline
Strong & 1.92 & 25.63 & 48.34 & 4.90 & 9.23 \tabularnewline
\hline
\end{tabular}
\end{table}

The effective mesh sizes of the three cases remain close, ranging from $1.64\times 10^{-3}$ to $1.92\times 10^{-3}~\mathrm{m}$. Therefore, the increase in the residual self-field cannot be explained simply by the change in local mesh size. Instead, it is mainly associated with grid stretching, which weakens the discrete compatibility among charge deposition, the Poisson gradient, and the gather operation.

The $K_x$ values remain below $10^{-20}$ even for the strong nonuniform
case. Thus, the self-field-induced displacement over one time step remains much
smaller than the local mesh size, although the residual self-field itself is
clearly amplified by grid stretching.

Additional tests with $r_{\min}=20000~\mathrm{m}$ give nearly identical results to those obtained with $r_{\min}=200~\mathrm{m}$. This confirms that $r_{\min}=200~\mathrm{m}$ is already sufficient to suppress the curvature effect and represent the Cartesian-limit behavior. The comparison therefore shows that, after the curvature effect is weakened, the residual self-field is governed mainly by the degree of grid nonuniformity rather than by the cylindrical coordinate curvature itself.

\section{Discussion}
\label{sec:discussion}

\subsection{Conservation, self-field cancellation, and residual scaling}
\label{subsec:discussion_conservation_self_field}

The verification tests show that the proposed deposition formulation is
consistent with the finite-volume geometry of nonuniform cylindrical
meshes. In the charge-deposition test, the use of nodal control volumes
removes the systematic density bias associated with the cylindrical
metric. In the current-deposition test, 
the swept-volume construction gives controlled current-density errors, while the
discrete continuity residual remains substantially below the current-density error.
These
results indicate that the scheme is conservative in the
charge-transport sense.

This property, however, is weaker than exact self-field cancellation.
Charge conservation constrains the integral charge balance and the time
evolution of the deposited density, whereas cancellation of the
numerical self-field requires a stronger compatibility among the charge
deposition, Poisson solve, discrete gradient, and particle gather. A
deposition scheme can therefore satisfy the discrete continuity equation
and still produce a finite self-field if the complete
scatter--solve--gradient--gather chain is not mutually compatible.

The self-field diagnostics confirm this distinction. Among the three
electric-field layouts tested on the same nonuniform cylindrical mesh,
the face-centered layout gives the smallest residual self-field, while
the cell-centered and shifted cell-centered layouts give larger
residuals. This result indicates that the residual is not controlled
only by the collocation of $\Phi$ and $\mathbf{E}$, but by the
compatibility of the full particle--field coupling chain. In the
present finite-volume formulation, the face-centered electric field is
more naturally paired with the cell-centered Poisson solve and the
discrete gradient operator.

The Cartesian-limit tests provide a complementary interpretation. Even
when the cylindrical curvature is weakened by shifting the radial
coordinate to a much larger radius, the residual self-field remains
sensitive to grid stretching. The default nonuniform grid gives a
self-field level comparable to that of the original cylindrical case,
whereas the uniform Cartesian-limit grid gives a much smaller residual.
Thus, within the tested Poisson--gradient--gather chains, grid
stretching is the main amplification mechanism, while cylindrical
curvature plays a secondary role.

The normalized one-step displacement $K_x$ provides a useful scale for
interpreting the residual self-field. In all cases tested here,
$K_x\ll 1$, so the displacement induced by the residual self-field over
one time step is much smaller than the local effective mesh size.
Nevertheless, because the residual varies systematically with field
placement and mesh stretching, weak numerical self-forces may still
accumulate in long-time simulations. Residual self-field diagnostics are
therefore useful when strongly stretched cylindrical meshes are used.

\subsection{Compatibility interpretation of the residual self-field}
\label{subsec:Compatibility interpretation of the residual self-field}

The present results are consistent with the general self-force behavior
reported for electrostatic PIC methods. In ideal structured Cartesian PIC,
a uniform grid with periodic boundaries and matched scatter--gather
operators can suppress the self-force to machine accuracy
\cite{hockney2021computer,lira2020self}. The present uniform
Cartesian-limit case is not identical to that ideal setting, because it is
computed in a finite cylindrical domain with zero Dirichlet boundaries.
Nevertheless, it gives the smallest residual level in this study.

For nonuniform meshes, finite boundaries, and unmatched operator chains, a
finite residual self-force is more difficult to avoid. The nonuniform
cylindrical cases studied here fall into this regime. Their RMS
self-fields are of order $10^{-5}~\mathrm{V/m}$ and remain below
$10^{-4}~\mathrm{V/m}$, while the corresponding normalized one-step
displacements remain in the range $10^{-21}$--$10^{-20}$. These values
are small in absolute displacement over one time step, but they are not
zero and they vary systematically with field placement and mesh
stretching.

This behavior also agrees with the compatibility viewpoint emphasized in
geometric PIC formulations on irregular or curvilinear grids. In such
methods, charge, current, fields, and metric information should be
represented by mutually compatible discrete objects, such as discrete
differential forms, Whitney interpolants, and Hodge operators
\cite{moon2015exact}. The present results show the same principle in a
finite-volume cylindrical setting: changing a single placement choice does
not guarantee a smaller self-force unless the full
scatter--solve--gradient--gather chain becomes more compatible.

\subsection{Implications and limitations}
\label{subsec:discussion_implications_limitations}

The first practical implication is that the face-centered electric-field
layout should be retained as the baseline layout for the present
finite-volume formulation. This layout is not favorable simply because
the field is stored on faces. It is favorable because it gives the most
compatible pairing among the tested cell-centered Poisson solve,
finite-volume gradient, and component-wise gather.

The second implication is that mesh stretching should be treated as a
numerical parameter that affects not only resolution, but also residual
self-field behavior. Nonuniform grids are useful for concentrating
resolution near regions of interest, but the Cartesian-limit tests show
that stretching can amplify the residual self-field even when the
curvature effect is weak. Therefore, residual self-field diagnostics,
such as $|E_{\mathrm{self}}|$ and $K_x$, should be included when
strongly stretched cylindrical meshes are used.


Several limitations should be noted. The self-field analysis is based on
a single-particle electrostatic diagnostic, which isolates the intrinsic
self-field of the discretization but does not include many-particle
noise, collective plasma response, magnetic-field coupling, or long-time
energy and momentum errors. The tests also use zero Dirichlet boundaries
and a simple logically rectangular cylindrical domain. More realistic
material boundaries, sheath-resolving geometries, and body-fitted meshes
may introduce additional compatibility errors. In addition, the present
formulation uses first-order particle shapes; higher-order shapes may
reduce local noise, but they also require correspondingly compatible
field reconstruction and gather operators.

\section{Conclusions}
\label{sec:conclusions}

This work examined charge and current density deposition on nonuniform
cylindrical PIC meshes and the residual numerical self-field generated by
the associated electrostatic particle--field coupling chain. A
cylindrical-volume-weighted deposition formulation was constructed using
nodal control volumes for charge deposition and swept-volume factors for
current deposition. The formulation accounts for both the cylindrical
metric and local mesh stretching, providing a finite-volume-consistent
basis for particle-to-grid mapping on stretched cylindrical grids.

The verification tests showed that the proposed formulation accurately
recovers a uniform charge density on both uniform and outer-refined
cylindrical grids. The controlled current-deposition test further showed
that the deposited current components 
reproduce the prescribed analytical currents with controlled errors, while the
discrete continuity residual remains substantially below the current-density error.
These results indicate that
the swept-volume current deposition is consistent with the charge variation
over one time step.

The self-field diagnostics showed that charge-transport consistency alone
does not guarantee exact self-field cancellation. Among the tested
field-placement layouts, the face-centered electric-field formulation
produced the smallest residual self-field, whereas the cell-centered and
shifted cell-centered layouts produced larger residuals. The
Cartesian-limit tests further showed that grid stretching remains a main
amplification mechanism even when cylindrical curvature is weakened. Thus,
the residual self-field is governed by the compatibility of the complete
scatter--solve--gradient--gather chain, rather than by charge conservation
or field collocation alone.

Overall, the proposed deposition formulation is charge-transport
consistent on the tested nonuniform cylindrical grids, but it is not
exactly self-force-free. In all cases considered, the normalized one-step
displacement $K_x$ remained far below unity, so the immediate displacement
caused by the residual self-field was much smaller than the local mesh
size. Nevertheless, the systematic dependence on field placement and grid
stretching suggests that residual self-field diagnostics should be included
when strongly nonuniform cylindrical meshes are used in long-time PIC
simulations. Future work should examine the scaling of the residual
self-field with mesh-stretching strength, particle-shape order, time-step
size, and boundary condition, and should explore more compatible
finite-volume PIC chains in which charge deposition, current deposition,
Poisson solve, discrete gradient, and particle gather are designed as a
unified set of operators.

\section*{Acknowledgment}

The authors acknowledge the support from National Natural Science Foundation of China (Grant No. 52472403).

\section*{Conflict of interest}
The authors have no conflicts to disclose.

\section*{Data Availability}

The data that support the findings of this study are available from
the corresponding author upon reasonable request.





\bibliography{reference}

@book{hockney2021computer,
  title     = {Computer Simulation Using Particles},
  author    = {Hockney, Roger W. and Eastwood, James W.},
  year      = {2021},
  publisher = {CRC Press},
  doi       = {10.1201/9780367806934},
  url       = {https://doi.org/10.1201/9780367806934}
}

@article{lira2020self,
  title     = {Self-force subtraction in particle in cell simulations},
  author    = {Resendiz Lira, Pedro Alberto and Marchand, Richard},
  journal   = {Computer Physics Communications},
  volume    = {254},
  pages     = {107212},
  year      = {2020},
  publisher = {Elsevier},
  doi       = {10.1016/j.cpc.2020.107212},
  url       = {https://doi.org/10.1016/j.cpc.2020.107212}
}

@article{moon2015exact,
  title     = {Exact charge-conserving scatter--gather algorithm for particle-in-cell simulations on unstructured grids: A geometric perspective},
  author    = {Moon, Haksu and Teixeira, Fernando L. and Omelchenko, Yuri A.},
  journal   = {Computer Physics Communications},
  volume    = {194},
  pages     = {43--53},
  year      = {2015},
  publisher = {Elsevier},
  doi       = {10.1016/j.cpc.2015.04.014},
  url       = {https://doi.org/10.1016/j.cpc.2015.04.014}
}

@article{villasenor1992rigorous,
  author  = {Villasenor, John and Buneman, Oscar},
  title   = {Rigorous charge conservation for local electromagnetic field solvers},
  journal = {Computer Physics Communications},
  volume  = {69},
  number  = {2--3},
  pages   = {306--316},
  year    = {1992},
  doi     = {10.1016/0010-4655(92)90169-Y}
}

@article{esirkepov2001exact,
  author  = {Esirkepov, T. Zh.},
  title   = {Exact charge conservation scheme for {Particle-in-Cell} simulation with an arbitrary form-factor},
  journal = {Computer Physics Communications},
  volume  = {135},
  number  = {2},
  pages   = {144--153},
  year    = {2001},
  doi     = {10.1016/S0010-4655(00)00228-9}
}

@article{ruyten1993density,
  author  = {Ruyten, Wilhelmus M.},
  title   = {Density-conserving shape factors for particle simulations in cylindrical and spherical coordinates},
  journal = {Journal of Computational Physics},
  volume  = {105},
  number  = {2},
  pages   = {224--232},
  year    = {1993},
  doi     = {10.1006/jcph.1993.1070}
}

@article{araki2014cell,
  author  = {Araki, Samuel J. and Wirz, Richard E.},
  title   = {Cell-centered particle weighting algorithm for {PIC} simulations in a non-uniform {2D} axisymmetric mesh},
  journal = {Journal of Computational Physics},
  volume  = {272},
  pages   = {218--226},
  year    = {2014},
  doi     = {10.1016/j.jcp.2014.04.037}
}

@article{verboncoeur2001symmetric,
  author  = {Verboncoeur, J. P.},
  title   = {Symmetric spline weighting for charge and current density in particle simulation},
  journal = {Journal of Computational Physics},
  volume  = {174},
  number  = {1},
  pages   = {421--427},
  year    = {2001},
  doi     = {10.1006/jcph.2001.6923}
}

@article{zhao2023rigorously,
  author  = {Zhao, Yinjian and Cui, Chen and Hu, Yuan},
  title   = {Rigorously conservative charge and current deposition in {3D} cylindrical {PIC}},
  journal = {Computational Particle Mechanics},
  volume  = {10},
  number  = {3},
  pages   = {495--502},
  year    = {2023},
  doi     = {10.1007/s40571-022-00513-6}
}

@article{stanier2022curvilinear,
  author  = {Stanier, A. and Chac{\'o}n, L.},
  title   = {A conservative implicit-{PIC} scheme for the hybrid kinetic-ion fluid-electron plasma model on curvilinear meshes},
  journal = {Journal of Computational Physics},
  volume  = {459},
  pages   = {111144},
  year    = {2022},
  doi     = {10.1016/j.jcp.2022.111144}
}

@article{zoni2022hybrid,
  author  = {Zoni, Edoardo and Lehe, R{\'e}mi and Shapoval, Olga and Belkin, Daniel and Za{\"i}m, Neil and Fedeli, Luca and Vincenti, Henri and Vay, Jean-Luc},
  title   = {A hybrid nodal-staggered pseudo-spectral electromagnetic particle-in-cell method with finite-order centering},
  journal = {Computer Physics Communications},
  volume  = {279},
  pages   = {108457},
  year    = {2022},
  doi     = {10.1016/j.cpc.2022.108457}
}

@article{shapoval2024pseudospectral,
  author  = {Shapoval, Olga and Zoni, Edoardo and Lehe, R{\'e}mi and Th{\'e}venet, Maxence and Vay, Jean-Luc},
  title   = {Pseudospectral particle-in-cell formulation with arbitrary charge and current-density time dependencies for the modeling of relativistic plasmas},
  journal = {Physical Review E},
  volume  = {110},
  number  = {2},
  pages   = {025206},
  year    = {2024},
  doi     = {10.1103/PhysRevE.110.025206}
}

@article{Chen2025PoP_PMSLPICHET3D,
author = {Chen, Xi and Xie, Lihuan and Zhong, Kunpeng and Luo, Xin and Zhou, Zhijun and Wang, Baisheng and Liu, Zhe and Zhao, Yinjian},
title = {Influence of Plume Region Arrangement on {Hall} Thruster Azimuthal Instability: {3D PIC} Simulations via a Newly Developed Code {PMSL-PIC-HET-3D}},
journal = {Physics of Plasmas},
volume = {32},
number = {5},
pages = {052103},
year = {2025},
doi = {10.1063/5.0253669}
}

@inproceedings{Zhao2025IEPC_MagneticFieldConfiguration,
  author       = {Zhao, Yinjian and Zhong, Kunpeng},
  title        = {Effect of Magnetic Field Configuration on {Hall} Thruster Azimuthal Instability in {3D PIC} Simulations},
  booktitle    = {Proceedings of the 39th International Electric Propulsion Conference},
  year         = {2025},
  address      = {London, United Kingdom},
  organization = {Electric Rocket Propulsion Society},
  note        = {IEPC-2025-063},
  url          = {https://www.electricrocket.org/IEPC_2025/London/submission_63.pdf}
}

@article{zhong2026effects,
author = {Zhong, Kunpeng and Zeng, Demai and Zhao, Yinjian and Yu, Daren},
title = {Effects of {RZ} Magnetic Field Components on Electron Drift Instability in {Hall} Thrusters via {3D PIC} Simulations},
journal = {Physics Letters A},
volume = {590},
pages = {131809},
year = {2026},
doi = {10.1016/j.physleta.2026.131809}
}

@inproceedings{chen2025iepcMCCNeutralSolver,
author = {Chen, Xi and others},
title = {Coupling {MCC} with Fluid Neutral Solver for {3D PIC} Simulation of {Hall} Thruster Azimuthal Instability},
booktitle = {Proceedings of the 39th International Electric Propulsion Conference},
year = {2025},
address = {London, United Kingdom},
organization = {Electric Rocket Propulsion Society},
note = {IEPC-2025-169},
url = {https://www.electricrocket.org/IEPC_2025/London/submission_169.pdf}
}

@article{li2005volume,
  author  = {Li, Yongdong and He, Feng and Liu, Chunliang},
  title   = {A volume-weighting cloud-in-cell model for particle simulation of axially symmetric plasmas},
  journal = {Plasma Science and Technology},
  volume  = {7},
  number  = {1},
  pages   = {2653--2656},
  year    = {2005},
  doi     = {10.1088/1009-0630/7/1/012}
}

@article{xiao2018structure,
  author  = {Xiao, Jianyuan and Qin, Hong and Liu, Jian},
  title   = {Structure-preserving geometric particle-in-cell methods for {Vlasov--Maxwell} systems},
  journal = {Plasma Science and Technology},
  volume  = {20},
  number  = {11},
  pages   = {110501},
  year    = {2018},
  doi     = {10.1088/2058-6272/aac3d1}
}

@article{xiao2021explicit,
  author  = {Xiao, Jianyuan and Qin, Hong},
  title   = {Explicit structure-preserving geometric particle-in-cell algorithm in curvilinear orthogonal coordinate systems and its applications to whole-device {6D} kinetic simulations of tokamak physics},
  journal = {Plasma Science and Technology},
  volume  = {23},
  number  = {5},
  pages   = {055102},
  year    = {2021},
  doi     = {10.1088/2058-6272/abf125}
}

@article{bao2025accelerated,
  author  = {Bao, Suxin and Liu, Jiyang and Zhu, Fujun and Liu, Gongzeng and Xing, Yan and Zhou, Zaifa},
  title   = {Accelerated field solver for {PIC/MCC} simulations via physics-informed neural networks},
  journal = {Plasma Science and Technology},
  volume  = {27},
  number  = {12},
  pages   = {125501},
  year    = {2025},
  doi     = {10.1088/2058-6272/ae055d}
}

@article{zhao2026review,
  author  = {Zhao, Zhongping and Zhao, Yinjian},
  title   = {A review of {3D} particle-in-cell simulations for electron drift instability in {Hall} thrusters},
  journal = {Plasma Science and Technology},
  volume  = {28},
  number  = {4},
  pages   = {045501},
  year    = {2026},
  doi     = {10.1088/2058-6272/ae69a9}
}

@article{taccogna2023plasma,
  title   = {Plasma propulsion modeling with particle-based algorithms},
  author  = {Taccogna, F. and Cichocki, F. and Eremin, D. and Fubiani, G. and Garrigues, L.},
  journal = {Journal of Applied Physics},
  volume  = {134},
  number  = {15},
  pages   = {150901},
  year    = {2023},
  doi     = {10.1063/5.0153862}
}

@article{adam2004study,
  title   = {Study of stationary plasma thrusters using two-dimensional fully kinetic simulations},
  author  = {Adam, J. C. and H{\'e}ron, A. and Laval, G.},
  journal = {Physics of Plasmas},
  volume  = {11},
  number  = {1},
  pages   = {295--305},
  year    = {2004},
  doi     = {10.1063/1.1632904}
}

@article{taccogna2005plasma,
  title   = {Plasma flow in a {Hall} thruster},
  author  = {Taccogna, Francesco and Longo, Savino and Capitelli, Mario and Schneider, Ralf},
  journal = {Physics of Plasmas},
  volume  = {12},
  number  = {4},
  pages   = {043502},
  year    = {2005},
  doi     = {10.1063/1.1862630}
}

@article{angus2024implicit,
  title   = {An implicit particle code with exact energy and charge conservation for studies of dense plasmas in axisymmetric geometries},
  author  = {Angus, Justin Ray and Farmer, William and Friedman, Alex and Geyko, Vasily and Ghosh, Debojyoti and Grote, Dave and Larson, David and Link, Anthony},
  journal = {Journal of Computational Physics},
  volume  = {519},
  pages   = {113427},
  year    = {2024},
  doi     = {10.1016/j.jcp.2024.113427}
}

@article{schmidt2012fully,
  title   = {Fully Kinetic Simulations of Dense Plasma Focus Z-Pinch Devices},
  author  = {Schmidt, A. and Tang, V. and Welch, D.},
  journal = {Physical Review Letters},
  volume  = {109},
  pages   = {205003},
  year    = {2012},
  doi     = {10.1103/PhysRevLett.109.205003}
}

@article{krishnan2012dense,
  title   = {The Dense Plasma Focus: A Versatile Dense Pinch for Diverse Applications},
  author  = {Krishnan, Mahadevan},
  journal = {IEEE Transactions on Plasma Science},
  volume  = {40},
  number  = {12},
  pages   = {3189--3221},
  year    = {2012},
  doi     = {10.1109/TPS.2012.2222676}
}

\end{document}